\documentclass[a4paper,11pt]{article}
\pdfoutput=1 

\usepackage{jheppub} 

\usepackage[T1]{fontenc} 
\usepackage{amssymb}
\usepackage{mathrsfs}
\usepackage{braket}
\usepackage{bbm}
\usepackage{hyperref}
\usepackage{cleveref}
\usepackage{transparent}
\usepackage{tikz}
\usepackage{bbm}

\title{\boldmath Large-c BCFT Entanglement Entropy with Deformed Boundaries from Emergent JT Gravity}

\author{Dominik Neuenfeld,}
\author{Christopher Tellinger}

\affiliation{University of Würzburg,\\Am Hubland, 97074 Würzburg, Germany}

\emailAdd{dominik.neuenfeld@uni-wuerzburg.de}
\emailAdd{christopher.tellinger@uni-wuerzburg.de}

\newtheorem{definition}{Definition}[section]

\abstract{
We study the effect of boundary deformations on the von Neumann entropy of subregions in two-dimensional boundary conformal field theories (BCFTs) at zero and finite temperature. The deformations considered are infinitesimal global conformal transformations that move the boundary and can equivalently be viewed as the leading-order effect of certain BCFT perturbations with the boundary displacement operator.

We demonstrate that at large central charge the von Neumann entropy in the presence of a deformed boundary is reproduced by the island entropy of the same interval in an undeformed BCFT acting as a bath and coupled to a gravitating spacetime. Here, the BCFT is joined to an AdS$_2$ region governed by Jackiw-Teitelboim (JT) gravity via transparent boundary conditions. The boundary condition for the dilaton field is set by the boundary deformation in the BCFT computation.

Our analysis relies on mild assumptions about the spectrum and OPE coefficients of the BCFT. Notably, these conditions are consistent with an exponentially large number of light operators and are therefore weaker than those required for holographic BCFTs. This is possible since the BCFT result is not reproduced by a gravitational computation in three dimensions with an $\mathcal O(1)$ number of matter fields, but instead by a computation in two dimensions and with an $\mathcal O(c)$ number of light fields. 
}

\begin{document} 
\maketitle
\flushbottom

\section{Introduction}
\label{sec:intro}
Two-dimensional gravity provides a uniquely tractable arena in which fundamental questions about quantum gravity can, in certain cases, be answered exactly. In recent years, Jackiw-Teitelboim (JT) gravity \cite{Jackiw:1984je,Teitelboim:1983ux,Almheiri:2014cka} and its boundary dual, the Schwarzian theory \cite{Almheiri:2014cka,Jensen:2016pah,Maldacena:2016upp,Engelsoy:2016xyb,Mertens:2018fds,Lam:2018pvp}, have served as the workhorse of this program. They arise as the universal low-energy effective descriptions of many systems, including the Sachdev-Ye-Kitaev model \cite{Sachdev:1992fk,Maldacena:2016hyu,Polchinski:2016xgd}, certain double-scaled matrix models \cite{Saad:2019lba,Stanford:2019vob,Suzuki:2021zbe}, the near-horizon dynamics of near-extremal black holes \cite{Strominger:1994tn,Maldacena:2016upp,Almheiri:2016fws,Nayak:2018qej,Moitra:2019bub}, and sectors of large-$c$ CFTs in certain limits of the left- and right-moving temperature \cite{Ghosh:2019rcj}. 

In this paper we provide evidence that JT gravity universally emerges in yet another system, namely as an effective description of infinitesimal deformations of the boundary of two-dimensional boundary conformal field theories (BCFTs) at large central charge.

Our work is motivated by bottom-up models of two-dimensional, holographic BCFTs with large boundary $g$-function \cite{Cardy:2004hm,AffleckLudwig1991}. In those systems the BCFT boundary is dual to a negatively curved end-of-the-world (ETW) brane embedded in an AdS$_3$ bulk \cite{Karch:2000ct,Takayanagi:2011zk,Almheiri:2019hni,Rozali:2019day}. An ETW brane is a dynamical co-dimension one hypersurface with Neumann (more precisely, umbilical) boundary conditions governed by an action which is usually taken to be a tension term. The brane intersects the asymptotic boundary of the locally AdS$_3$ bulk at the location of the BCFT boundary and geometrizes the boundary degrees of freedom. In \cite{Neuenfeld:2024gta} it was argued using perturbation theory that the effective theory which describes displacements of the ETW brane is given by Liouville theory with a dynamical background metric. Using a $T\bar T$-perspective, the authors of \cite{Callebaut:2025thw} demonstrated that at least for Randall-Sundrum models \cite{Randall:1999vf} this statement holds to all orders. Restricting to linearized fluctuations, the effective theory of brane displacements simplifies from Liouville theory to JT gravity \cite{Geng:2022slq,Aguilar-Gutierrez:2023tic,Neuenfeld:2024gta}. In both cases, turning on a non-trivial dilaton changes the location of the brane near the asymptotic boundary and thus amounts to moving the BCFT boundary.

This relation between a BCFT$_2$ and an ETW brane in AdS$_3$ can be turned into a purely two-dimensional statement between a BCFT$_2$ and a CFT$_2$ coupled to a gravitating region by integrating out the bulk while keeping the asymptotic boundary as well as the brane \cite{Suzuki:2022xwv}.\footnote{The resulting description is known as the intermediate or brane-description in double-holographic systems \cite{Almheiri:2019hni,Geng:2020qvw,Chen:2020uac}.} This produces a description of the system in which a CFT$_2$ lives on a spacetime given by the asymptotic boundary glued onto the AdS$_2$ brane worldvolume. The part of the background geometry which arises from the brane is coupled to JT gravity and can exchange matter excitations with the non-gravitating part of the spacetime, dubbed the bath. The gravitational coupling constant is related to the central charge as
\begin{align}
\label{eq:gnewton}
    c = \frac{3}{2G_\mathrm{N}}.
\end{align}
Note that this is not the Brown-Henneaux relation \cite{Brown:1986nw} between the gravitational coupling in three-dimensional AdS space and the central charge of a dual theory. Instead, it differs by a factor of the AdS radius and \cref{eq:gnewton} relates the gravitational constant in a two-dimensional theory to the central charge of a two-dimensional CFT.
The results of \cite{Neuenfeld:2024gta} then turn into the claim that two-dimensional holographic BCFTs with large boundary $g$-function have an alternative description in which the boundary is replaced by a region of dynamical JT gravity which CFT excitations can freely enter and leave.

To be more precise about the relation between the dilaton and boundary deformations, it was argued in \cite{Neuenfeld:2024gta} that turning on a non-trivial JT gravity solution on the brane amounts to an infinitesimal displacement of the BCFT boundary by an amount proportional to the renormalized dilaton in the dual description. Let us illustrate this by considering a holographic BCFT$_2$ on a half-space, dual to the  Poincar\'e patch of AdS$_3$ which is cut off by a two-dimensional ETW brane $\mathcal Q_0$. According to our discussion above, an infinitesimally displaced brane $\mathcal Q$ can be described as a non-trivial dilaton solution of JT gravity on the worldvolume of $\mathcal Q_0$. The brane displacement, or equivalently the dilaton, diverges along the brane as $\phi(\tau,y) \sim \frac {f(\tau)} y$ near the boundary at $y = 0$. This leads to a deformation of the BCFT$_2$ boundary proportional to $\delta X(\tau) \sim f(\tau)$.
A naive application of the extrapolate dictionary to the AdS$_2$ brane,
\begin{align}
    f(\tau) = \lim_{y \to 0} y^{-\Delta_-} \phi(\tau, y),
\end{align}
with $\Delta_- = -1$, then suggests that the asymptotic value of the dilaton sources a BCFT operator with dimension $\Delta_+ = 2$. Since the dilaton lives on the brane, this operator must moreover be located at the BCFT boundary. And in fact in BCFTs there is a universal operator with these properties:\footnote{More generally, this operator exists in any theory with boundaries or defects. For example in interface CFTs the normalization of its two-point function encodes physical data about the transmission properties of the interface \cite{Meineri:2019ycm}.} the displacement operator $\mathcal D(\tau) = T^{nn}(\tau)$ which describes the response to a displacement of the boundary \cite{Billo:2016cpy} and is given by the normal-normal component of the stress-energy tensor evaluated at the boundary. Phrased differently, in the case of holographic BCFTs, an irrelevant deformation of the boundary with the displacement operator can equivalently be described as coupling the system to a new AdS$_2$ region with JT gravity, where the asymptotic behavior of the dilaton is determined by the BCFT boundary deformation.

It would certainly be interesting to demonstrate the above statements directly from a BCFT computation. Moreover, one might wonder which properties a BCFT needs such that JT gravity arises universally as an effective description of a deformed boundary.

In this paper we will make progress on answering this question by studying entanglement entropies in BCFTs. We consider two-dimensional, large-$c$ BCFTs at leading order in the central charge on the plane or the cylinder with boundary at $\textrm{Im}(z) = 0$. We deform the boundary with an infinitesimal global conformal transformation
\begin{align}
    z \to z - \mathrm{i} \Delta X(z)
\end{align}
and demonstrate that the entanglement entropy in this deformed background agrees with the island entropy of a CFT in an AdS$_2$/bath system coupled to JT gravity if the dilaton solution satisfies
\begin{align}
    \label{eq:deformation_expectation}
    \Delta X(\tau) = \lim_{y \to 0} y\, \phi(\tau,y).
\end{align}
Additionally, we show that the coefficient $\phi_0$ of the topological term of JT gravity is related to the boundary entropy $\log g_b$ of the BCFT, such that both measure the ground state degeneracy.

In our work we go beyond the analyses in \cite{Geng:2022slq,Aguilar-Gutierrez:2023tic,Neuenfeld:2024gta,Callebaut:2025thw} which were performed from the gravitational point of view in AdS$_3$, i.e., assuming a holographic BCFT as well as the validity of modeling the dual of the BCFT boundary as an ETW brane. Instead, we will approach the problem from a purely CFT perspective by computing and comparing subregion entropies in two different systems.

The first system will simply be called the \emph{BCFT system}. In the simplest example we consider, it consists of a two-dimensional BCFT on the complex upper half-plane with its boundary infinitesimally deformed by a broken global conformal transformation. We consider subsystems of the BCFT and compute their von Neumann entropy using the replica trick \cite{Calabrese:2009qy}. The result is the von Neumann entropy of a BCFT subregion on the upper half-plane plus a correction term which depends on the deformation. We then compare this result to an entanglement computation in a second system.

\begin{figure}[t]
    \centering
\def\svgwidth{14cm}
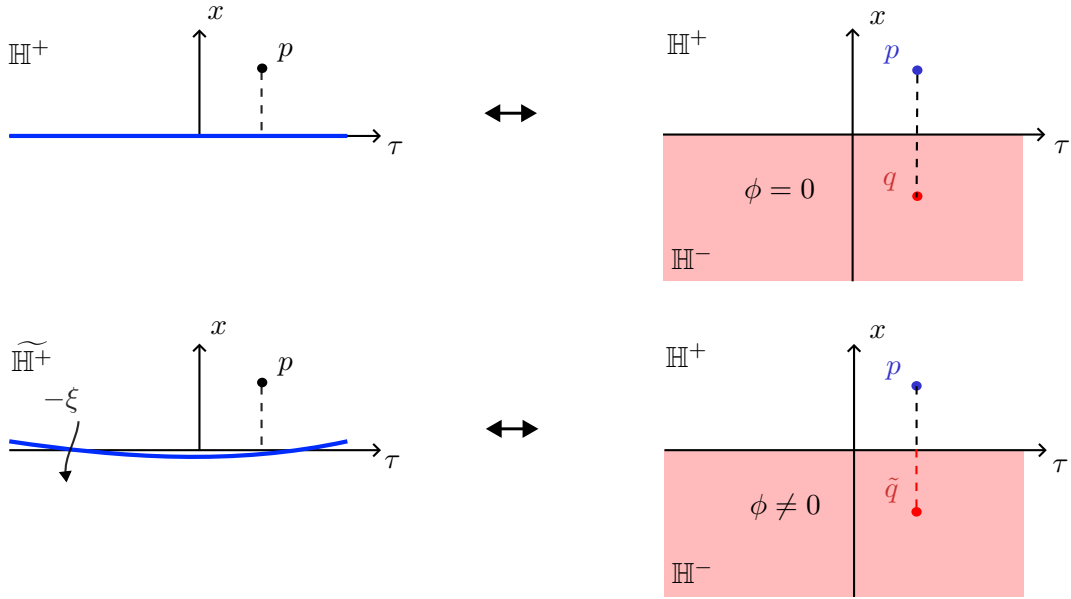
\caption{The von Neumann entropy in a BCFT$_2$ can be mapped to the island entropy in an AdS$_2$/bath system. Deformations of the BCFT boundary map, at leading order, to turning on a non-trivial dilaton in the gravitational description.}
\label{fig:dual}
\end{figure}

We call the second system the \emph{AdS$_2$/bath system}. It consists of a two-dimensional CFT which lives on a space partially coupled to gravity. In the simplest example, the CFT lives on the complex plane. The upper half-plane acts as a non-gravitating bath and we equip the lower half plane with an AdS metric and couple it to JT gravity. 
Such models have played an important role in studying the Page curve of black hole evaporation \cite{Almheiri:2019hni,Geng:2020qvw,Chen:2020uac}. We consider a dilaton profile that is set by the boundary deformation in the first system as explained earlier and choose transparent boundary conditions between the two half-planes such that CFT excitations can cross freely between the gravitating and non-gravitating region. In this setup we consider the same interval as in the BCFT system, but compute its entropy using the quantum extremal island formula \cite{Almheiri:2019psf,Penington:2019npb}, which is the natural notion of entropy is this setting. Both systems are illustrated in \cref{fig:dual}.

As we will see, the von Neumann entropies of subregions in the BCFT system agree with the island entropy in the AdS$_2$/bath system. Remarkably --- at least for the case of entanglement entropies under consideration --- the BCFT need not be holographic. Rather, weaker assumptions on the spectrum, which we will call weak vacuum block dominance, are sufficient to arrive at our result. In particular, while we need to require that for more than a single interval the functional form of correlators is determined by the vacuum block at large $c$, the coefficient of the block can be exponentially large in $c$. This happens for example when the CFT has exponentially many light operators. Thus our results also apply to certain non-holographic CFTs.

The remainder of this paper is organized as follows. We review the necessary aspects of BCFTs as well as JT gravity in \cref{sec:review}. \Cref{sec:single_interval} studies the behavior of subregion entropy of a single interval. Since results for a single interval are completely determined by conformal symmetry we study the case of two intervals in \cref{sec:multiple_intervals}. There, we also discuss the detailed constraints we need to impose on the spectrum and OPE coefficients of the BCFT such that an AdS$_2$/bath description can be used to compute entropies. The generalization to an arbitrary number of intervals is discussed in \cref{sec:multiple_intervals_2}. We conclude with further directions in \cref{sec:discussion}.

\section{Review}
\label{sec:review}
In this section, we will review the necessary background material and set up our conventions. Readers familiar with BCFTs, JT gravity, and the quantum extremal island formula are invited to skip this section and refer to it should the need arise. 

In the following, we will review the basics of BCFTs and the conformal block expansion and how entanglement entropies are computed in BCFT$_2$ using the replica trick. We then give a lightning review of JT. Lastly, we briefly discuss how in the presence of gravity, the natural notion of subregion entropy is computed by the island formula. Throughout this work we will be working in Euclidean signature on the complex plane with coordinates $z, \bar z$ and $z =\tau+\mathrm{i}x$. We will mostly consider BCFTs living on the upper half-plane $\mathbb H^+$, such that the boundary is the real line parametrized by $\tau$.

\subsection{Boundary Conformal Field Theory}
\label{sec:BCFT}
A BCFT$_2$ is a two-dimensional conformal field theory on a space with boundary, such that the boundary conditions preserve half of the conformal symmetry.

\subsubsection*{Boundary States and Boundary Operators}

The presence of a boundary leads to new physical data, encoded in boundary conditions and boundary-localized operators. A two-dimensional CFT without boundaries has left- and right-moving Virasoro symmetries, which are generated by $L_n,\overline{L}_n$. These generators satisfy the Virasoro algebra 
\begin{equation}
    [L_n, L_m]=(n-m)L_{n+m}+\frac{c}{12}n(n^2-1)\delta_{n+m,0}
\end{equation} 
and similarly for $\overline{L}_n$.

Conformal invariance in the presence of a boundary requires the identification of left- and right-moving components of the stress tensor at the boundary. In Euclidean coordinates with boundary at $z=\bar z$, this condition reads \begin{equation}
\label{eq:Tbdrycond}
    T(z)=\overline{T}(\overline{z})
\end{equation} at the boundary. Boundary conditions satisfying this constraint define conformal boundary conditions. This in particular implies that the $\text{Vir} \times \overline{\text{Vir}}$ symmetry algebra is broken to a diagonal subalgebra. 

A powerful way to encode these boundary conditions is through boundary states. Upon mapping the theory to the closed-string channel, a conformal boundary condition $b$ is (formally) represented as a state $\ket{b}$ of the bulk CFT.\footnote{Boundary states are not normalizable and thus strictly speaking do not live in the CFT Hilbert space. However, they can be made normalizable by evolving in Euclidean time by some finite amount and thus correlation functions in a boundary state are well-defined.} Conformal invariance is then realized as the condition
\begin{equation}
    (L_n-\overline{L}_{-n})\ket{b}=0
\end{equation} for all $n\in\mathbb{Z}$, see \cite{Cardy:2004hm, blumenhagen2012basic}. 

One can compute the overlap of $\ket{b}$ with the vacuum state $\ket{0}$ by calculating the 
disk amplitude \begin{equation}
    g_b=\braket{0|b}. \label{diskpart}
\end{equation} 
This amplitude is called the $g$-function associated with the conformal boundary condition $b$. In quantum impurity problems, $g_b$ counts the effective ground-state degeneracy associated with the boundary degrees of freedom \cite{AffleckLudwig1991}. Under boundary renormalization group flows, the $g$-function is monotonically decreasing and thus for the boundary it plays a role similar to the central charge for a CFT$_2$. Below, we will recall how it appears as a boundary contribution to entanglement entropy for subregions that contain part of the boundary.

A crucial ingredient in the definition of a CFT are the coefficients $\mathscr{C}_{ij}^k$ which appear in the operator product expansion (OPE) of two operators\footnote{Here, we chose scalar operators for simplicity.}
\begin{equation}
    \mathcal{O}_i(z_1, \bar{z}_1)\mathcal{O}_j(z_2, \bar{z}_2) = \sum_k \frac{\mathscr{C}_{ij}^k}{|z_1-z_2|^{\Delta_i+\Delta_j - \Delta_k}} \mathcal C[z_1 - z_2, \partial_z]\mathcal{O}_k(z_2)
\end{equation} with the respective operator dimensions $\Delta_i, \Delta_j, \Delta_k$, see e.g.\ \cite{schottenloher1997mathematical}. Here, the sum runs over bulk CFT primary operators and $\mathcal C$ is a differential operator which depends on the conformal dimensions of the involved operators and is normalized such that its term at zeroth order in derivatives is simply $1$.

In a BCFT the presence of a boundary moreover allows for the existence of a special set of operators localized at the boundary. Such operators $\hat{\mathcal{O}}_J$, which we will denote with a hat, come with a conformal scaling dimension $\hat{\Delta}_J$.
Boundary operators arise naturally in the operator expansion of bulk operators approaching the boundary (BOE). For a bulk primary operator $\mathcal{O}_i$, the BOE takes the form \begin{equation}
    \label{eq:bulk_boundary_ope}
    \mathcal{O}_i(z,\bar{z})=\sum_J \frac{\mathscr{B}_{iJ}^{b}}{(2x)^{\Delta_i-\hat{\Delta}_J}}\tilde{\mathcal{C}}[x,\partial_\tau]\hat{\mathcal O}_J(\tau),
\end{equation} 
where $\Delta_i=h_i+\overline{h}_i$ is the bulk scaling dimension, and $\mathscr{B}^b_{iJ}$ are the BOE coefficients. We will use the notation $\mathscr{B}^b_{iJ}$ and $\mathscr{B}_{\mathcal O_i \hat {\mathcal O}_J}^{b}$ interchangeably. The index $J$ runs over all boundary primary operators and $\tilde{\mathcal{C}}$ is a differential operator which depends only on the conformal weights of $\mathcal{O}_i$ and $\hat{\mathcal{O}}_J$. From now on we will leave the dependence of OPE coefficients and correlation functions on the boundary condition $b$ implicit.

Analogously to the OPE expansion, we have an OPE of two boundary operators in a BCFT, the boundary OPE (BOPE) of the form
\begin{equation}
\label{eq:BOPE}
    \hat{\mathcal{O}}_I(\tau)\hat{\mathcal{O}}_J(\tau')=\sum_K\frac{\mathscr{D}_{IJK}}{|\tau-\tau'|^{\hat{\Delta}_I+\hat{\Delta}_J- \hat{\Delta}_K}}\tilde{ \tilde{\mathcal C}}[\tau-\tau', \partial_{\tau'}]\hat{\mathcal{O}}_K(\tau').
\end{equation}

\subsubsection*{BCFT Correlation Functions}
In a CFT the form of correlation functions is strongly constrained by conformal symmetry.

Correlation functions in a BCFT are less constrained, since the conformal symmetry is partially broken. The doubling trick \cite{Cardy:2004hm, blumenhagen2012basic} tells us that the form of a BCFT correlator on $\mathbb{H}^+$ with $n$ operators of dimensions $(h_i, \bar h_i)$, inserted at $z_1, \dots, z_n$ is the same as that of a chiral CFT on $\mathbb{C}$ with $2n$ operators inserted at $z_1, \dots, z_n$ and mirror points $z^*_1=\bar{z}_1, \dots, z^*_n=\bar{z}_n$ with dimensions $h_i$ and $\bar h_i$, respectively. 

An immediate consequence is that, unlike in a CFT, scalar primaries in a BCFT are allowed to have non-vanishing one-point functions which are functions of the distance to the boundary,
\begin{equation}
    \braket{\mathcal{O}_{\Delta}(z,\bar{z})}_{\mathbb{H}^+}=\frac{\mathscr{A}_\mathcal{O}}{|z-\bar{z}|^\Delta}. \label{1P}
\end{equation} The coefficient $\mathscr{A}_\mathcal{O}$ depends on the boundary condition $b$ and the inserted operator $\mathcal{O}$.

Inserting a bulk operator $\mathcal{O}_i$ at $z\in\mathbb{H}^+$ and a boundary operator $\hat{\mathcal{O}}_I$ at $\tau'\in\mathbb{R}$, one can write down a bulk-boundary-operator two-point function
\begin{equation}
\braket{\mathcal{O}_i(z,\bar{z})\hat{\mathcal{O}}_I(\tau')}_{\mathbb{H}^+}=\frac{\mathscr{B}_{\mathcal{O}_i\hat{\mathcal{O}}_I}}{(2x)^{\Delta_i-\hat{\Delta}_I}(x^2+(\tau-\tau')^2)^{\hat{\Delta}_I}}
\end{equation} where, as before, $z= \tau + i x$. Comparing this equation with \cref{1P} it is clear that for the identity operator $\hat{\mathcal{O}}_I= \mathbbm{1}$, we have $\mathscr{B}_{\mathcal{O}_i\mathbbm{1}}=\mathscr{A}_{\mathcal{O}_i}$.

\begin{figure}[t]
    \centering
\def\svgwidth{8cm}
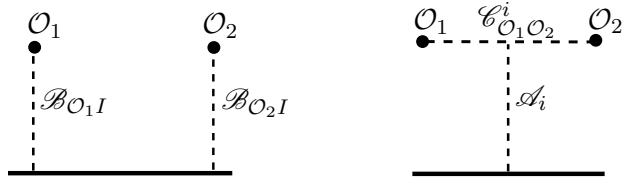
\caption{Visualization of the boundary (left) and the bulk channel (right).}
\label{fig:channels}
\end{figure}

A BCFT bulk two-point function of two bulk primary operators $\mathcal{O}_1,\mathcal{O}_2$ with equal scaling dimension $\Delta$ is only determined up to a function $F(\eta)$,
\begin{equation}
\braket{\mathcal{O}_1(z_1,\bar{z}_1)\mathcal{O}_2(z_2,\bar{z}_2)}_{\mathbb{H}^+}=\left(\frac{\eta}{4x_1 x_2}\right)^\Delta F(\eta), \label{4point}
\end{equation} where the BCFT conformal cross-ratio $\eta$ of the points $z_1, z_2$ is defined by
\begin{equation}
    \eta =\frac{(z_1-z^*_1)(z_2-z^*_2)}{(z_1-z^*_2)(z_2-z^*_1)}.
\end{equation} 
Note that unlike the conformal cross-ratio in a two-dimensional CFT
\begin{align}
    \eta^\text{CFT}=\frac{(z_1-z_2)(z_3-z_4)}{(z_1-z_3)(z_2-z_4)}
\end{align}
which is constructed from four independent points, the BCFT cross-ratio is constructed from only two points and their complex conjugates. In the following we will use $\eta$ to denote BCFT and CFT cross-ratios. It will be clear from the context which one we are referring to. For later considerations, the following relation for BCFT $\eta$ will become important: \begin{equation}
    \frac{\eta}{4x_1 x_2}=\frac{1-\eta}{|z_1-z_2|^2}. \label{crossRel}
\end{equation} 
The function $F(\eta)$ in \eqref{4point} is a function of the BCFT conformal cross-ratio $\eta$ only and does not explicitly depend on $z_1$ and $z_2$. It can be expressed in two equivalent ways, see \cite{Sully:2020pza} for a nice review:
\begin{enumerate}
    \item We can expand each bulk operator using the BOE in boundary operators, \cref{eq:bulk_boundary_ope}, resulting in a sum over boundary conformal blocks controlled by bulk-boundary and boundary OPE data. In this case, we have \begin{equation}
    F(\eta)=\sum_I\mathscr{B}_{\mathcal{O}_1I} \mathscr{B}_{\mathcal{O}_2I}\mathscr{F}(c,\hat{\Delta}_I,\Delta/2\,|\,\eta).
\end{equation} The index $I$ labels the boundary primary operators, $c$ is the central charge, and $\hat{\Delta}_I$ denotes the scaling dimension of the internal boundary operators. $\mathscr{F}$ is a chiral \emph{conformal block} and depends, apart from the previously mentioned quantities, only on the cross-ratio $\eta$.
\item We can take the bulk OPE between the two bulk operators. This leads to a sum over conformal blocks weighted by bulk OPE coefficients $\mathscr{C}_{\mathcal{O}_1\mathcal{O}_2}^{i}$ and one point function coefficients. In this expansion, we have
\begin{equation}
    F(\eta)=\sum_i\mathscr{C}_{\mathcal{O}_1\mathcal{O}_2}^{i}\mathscr{A}_i\mathscr{F}(c,\Delta_i,\Delta/2\,|\,1-\eta),
\end{equation} 
with the same conformal block $\mathscr{F}$.
\end{enumerate} 
Crossing symmetry requires the equality of these two decompositions, providing powerful consistency conditions on BCFT data. The visualization of the two channels is shown in \cref{fig:channels}, where the two bulk operators $\mathcal{O}_1$ and $\mathcal{O}_2$ are evaluated at the points $z_1$ and $z_2$, respectively.

\subsubsection*{Entanglement Entropy}
Even in the vacuum state, the degrees of freedom of a local quantum field theory located in different subregions are entangled. One way of quantifying this entanglement is the von Neumann entropy. Consider a system in a state described by the density matrix $\rho$. We can split the system into two disjoint subregions $A$ and $\bar A$.
The reduced density matrix of the subsystem $A$ can be obtained by tracing out the complement, i.e. \begin{equation}
    \rho_A=\mathrm{tr}_{\bar A}\rho.
\end{equation} The von Neumann entropy, or entanglement entropy, of $A$ is then given by \begin{equation}
    S(A)=-\mathrm{tr}(\rho_A\log\rho_A).
\end{equation} With the definition of the $n$-th R\'enyi entropy \begin{equation}
    S^{(n)}(A)=\frac{1}{1-n}\log\mathrm{tr}(\rho_A^n), \label{Renyi}
\end{equation} one can obtain the von Neumann entropy by taking the analytic continuation \begin{equation}
    S(A)=\lim_{n\to 1} S^{(n)}(A). \label{entdef}
\end{equation}
In $2d$ CFTs or BCFTs on a manifold $\mathcal M$ there exists an elegant method for computing R\'enyi entropies and therefore also entanglement entropies: the replica approach \cite{Calabrese:2004eu, Cardy:2007mb, Calabrese:2009qy}. To compute the von Neumann entropy of the interval $A = [z_1, z_2]$ we cut the manifold open along $A$ and cyclically glue $n$ copies of $\mathcal M$ across their cuts. The resulting manifold $\mathcal{R}_n$ is then a branched $n$-fold cover of the original background geometry.

The replica trick allows us to compute $\mathrm{tr}\,\rho_A^n$ via \begin{equation}
    \mathrm{tr}\,\rho_A^n=\frac{Z_n}{Z_1^n}
\end{equation} where $Z_n$ is the CFT partition function on the replica geometry $\mathcal{R}_n$ and $Z_1$ the partition function on the original geometry $\mathcal R _1$. We can then apply \cref{entdef} to obtain the von Neumann entropy. Due to UV divergences near the boundary of $A$, the resulting expression diverges and needs to be regulated. We will discuss how this is done below.

In practice, the ratio $\frac{Z_n}{Z_1^n}$ is computed as a correlation function of twist fields inserted at the endpoints of $A$ in the CFT. Formally, a twist operator $\Phi_n$ is a scalar primary operator with scaling dimension \begin{equation}
    \Delta=d_n=\frac{c}{12}\left(n-\frac{1}{n}\right).
\end{equation} 
In a CFT, twist operators must appear in conjugate pairs $\Phi_n, \overline{\Phi}_n$ with an associated branch cut connecting the two insertions. Encircling a twist insertion moves between different replica sheets.

\subsubsection*{BCFT Entanglement Entropy} \label{subsec Ent}
\begin{figure}[!t]
    \centering
\def\svgwidth{5cm}
\begingroup%
  \makeatletter%
  \providecommand\color[2][]{%
    \errmessage{(Inkscape) Color is used for the text in Inkscape, but the package 'color.sty' is not loaded}%
    \renewcommand\color[2][]{}%
  }%
  \providecommand\transparent[1]{%
    \errmessage{(Inkscape) Transparency is used (non-zero) for the text in Inkscape, but the package 'transparent.sty' is not loaded}%
    \renewcommand\transparent[1]{}%
  }%
  \providecommand\rotatebox[2]{#2}%
  \newcommand*\fsize{\dimexpr\f@size pt\relax}%
  \newcommand*\lineheight[1]{\fontsize{\fsize}{#1\fsize}\selectfont}%
  \ifx\svgwidth\undefined%
    \setlength{\unitlength}{184.2519685bp}%
    \ifx\svgscale\undefined%
      \relax%
    \else%
      \setlength{\unitlength}{\unitlength * \real{\svgscale}}%
    \fi%
  \else%
    \setlength{\unitlength}{\svgwidth}%
  \fi%
  \global\let\svgwidth\undefined%
  \global\let\svgscale\undefined%
  \makeatother%
  \begin{picture}(1,0.36097433)%
    \lineheight{1}%
    \setlength\tabcolsep{0pt}%
    \put(0,0){\includegraphics[width=\unitlength,page=1]{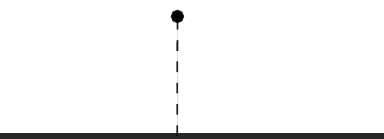}}%
    \put(0.51049805,0.321078){\color[rgb]{0,0,0}\makebox(0,0)[lt]{\lineheight{1.25}\smash{\begin{tabular}[t]{l}$z$\end{tabular}}}}%
    \put(0.59002909,0.17921181){\color[rgb]{0,0,0}\transparent{0.85156202}\makebox(0,0)[lt]{\lineheight{1.25}\smash{\begin{tabular}[t]{l}$A$\end{tabular}}}}%
    \put(0,0){\includegraphics[width=\unitlength,page=2]{Interval.pdf}}%
    \put(0.01489571,0.28732125){\color[rgb]{0,0,0}\makebox(0,0)[lt]{\lineheight{1.25}\smash{\begin{tabular}[t]{l}$\mathbb{H}^+$\end{tabular}}}}%
  \end{picture}%
\endgroup%

\caption{An interval $A$ which ends on the boundary of a BCFT.}
\label{fig:interval}
\end{figure}

The computation of entanglement entropies for subregions in BCFTs works almost analogously to the CFT case. However, a key difference is that in a BCFT an isolated twist operator is allowed with its associated cut terminating on the boundary, see \cref{fig:interval}.
As a starting point to compute the entanglement entropy of a single interval $A$ from a point $z\in\mathbb{H}^+$ to the boundary we have \begin{equation}
    \mathrm{tr}\,\rho_A^n=\frac{Z_n}{Z_1^n}
    =\braket{\Phi_n(z,\overline{z})}_{\mathbb{H}^+}
    =\mathscr{A}_{\Phi_n}|z-\bar{z}|^{-d_n}, \label{Twistonepoint}
\end{equation} where we have used \cref{1P}.
Next, we can use \cref{Renyi} to obtain the R\'enyi entropy \begin{equation}
    S^{(n)}(A)=\frac{c}{12}\frac{n+1}{n}\log|z-\bar{z}|+\frac{1}{1-n}\log\mathscr{A}_{\Phi_n}. \label{2DRenyi}
\end{equation}
Applying \cref{entdef} to obtain the entanglement entropy is subtle, however, since it depends on how $\mathscr{A}_{\Phi_n}$ behaves as $n \to 1$. It is therefore necessary to find an explicit expressions for $\mathscr{A}_{\Phi_n}$ which also happens to capture the regulator dependence advertised earlier \cite{Ohmori_2015, Sully:2020pza}.

To find such an explicit expression consider the conformal map 
\begin{equation}
    \label{eq:conf_map_cyl}
    k: w\mapsto\mathrm{i}\log\frac{w-\bar{z}}{w-z},
\end{equation}
which maps the upper half-plane to a semi-infinite cylinder. More precisely, under this map the real line $w \in \mathbb R$, i.e., the boundary, is compactified to $\mathrm{Im}(k(w)) = 0$, $\mathrm{Re}(k(w)) = [0,2\pi)$. The point $z$, i.e., the endpoint of the interval $A$ is mapped to infinity. This infinite distance can be regulated by placing a boundary with boundary condition $a$ at a small circle with radius $\epsilon$ around $z$. Under the map $k$ this regulator surface maps to 
$\mathrm{Im}\,k(w)=\log\frac{2 x}{\epsilon}$
with $x=\mathrm{Im}(z)$ for $\epsilon\ll1$ and cuts off the cylinder at a finite distance. Thus, \cref{eq:conf_map_cyl} maps our configuration onto a cylinder with length $\tau=\log\frac{2 x}{\epsilon}$ with two boundaries and boundary conditions $\mathrm{a}$ and $\mathrm{b}$ on either end.
The replica geometry $\mathcal{R}_n$ is defined by cutting open the cylinder along the cylinder axis and gluing $n$ of the resulting sheets together. Using boundary states $\ket{a},\ket{b}$, we can then write the partition function on this surface as
\begin{equation}
    Z_n=\big\langle a|\mathrm{e}^{-\frac{\tau}{2\pi n}H}|b\big\rangle
\end{equation}
with the CFT Hamiltonian $H$. For large Euclidean time $\tau$, time evolution projects onto the vacuum
    \begin{equation}
    \mathrm{e}^{-\frac{\tau}{2\pi n}H}\approx \mathrm{e}^{-\frac{\tau}{2\pi n}E_0}\ket{0}\bra{0}
\end{equation} and with the vacuum energy $E_0=-\frac{\pi c}{6}$ for a CFT on a circle, we obtain in the limit $\epsilon\to 0$ \begin{equation}
    \braket{\Phi_n(z,\overline{z})}_{\mathbb{H}^+}=(\braket{a|0}\braket{0|b})^{(1-n)}\left(\frac{|z-\bar{z}|}{\epsilon}\right)^{-d_n}.
\end{equation} 
Comparing with \cref{Twistonepoint} we can read off 
\begin{equation}
    \mathscr{A}_{\Phi_n}=(\braket{a|0}\braket{0|b})^{(1-n)}\epsilon^{d_n}, \label{A Def}
\end{equation} and finally use \cref{Renyi,entdef} to find \begin{equation}
    S(A)=\frac{c}{6}\log\frac{|z-\bar{z}|}{\epsilon}+\log g_a+\log g_b.
\end{equation} 
Here, we have used the definition of the $g$-function, \cref{diskpart}, for the boundary conditions $a$ and $b$. The boundary condition $a$ is arbitrary (up to the fact that it needs a non-vanishing overlap with the vacuum) and is associated with the regulator $\epsilon$. Therefore, we can redefine $\epsilon$ to absorb the term $\log g_a$. However, $\log g_b$ is associated to the boundary of our BCFT and, as a physical boundary entropy, can not be defined away. Finally, we find the well-known result \begin{equation}
S(A)=\frac{c}{6}\log\frac{|z-\bar{z}|}{\epsilon}+\log g_b=\frac{\mathrm{c}}{6}\log\frac{2x}{\epsilon}+\log g_b. \label{EntSingle}
\end{equation} 

\subsection{AdS/Bath Geometry and Island Entropy}
\label{Subsec AdS$_2$/Bath Geometry and generalized Entropy}
We now turn to an a priori unrelated setup in which a CFT$_2$ lives on a spacetime which consists of a gravitating region and a non-gravitating \emph{reservoir} or \emph{bath}. Systems in which gravitating regions are allowed to exchange excitations with non-gravitating regions have been in the focus of interest recently. For example, they allow to derive the Page-curve for black hole evaporation using the quantum extremal surface (QES) prescription for entanglement entropy \cite{Almheiri:2019psf,Penington:2019npb,Rozali:2019day}, providing an explicit demonstration of how information is encoded in Hawking radiation radiation as a black hole evaporates. More generally, they are well-controlled gravitational laboratories to study the effect of gravity in open systems, see e.g. \cite{Geng:2020fxl,Karch:2025hof}.
Among the most studied examples is the coupling of a region with two-dimensional Jackiw–Teitelboim gravity to a conformal bath \cite{Almheiri:2019hni}, which will also be our focus. 

\subsubsection*{AdS/Bath Systems} 
We consider a CFT$_2$ living on the complex plane which we divide into the upper and lower half-plane. The upper half-plane $\mathbb H^+$ is equipped with the Euclidean metric and acts as a bath region. On the lower half-plane $\mathbb H^-$ we introduce a dynamical metric $g_{\mu\nu}$ and a dilaton $\phi$. On $\mathbb H^-$ the CFT is minimally coupled to the metric, but not to the dilaton. Excitations are free to cross between the upper and lower half-plane and can carry energy and information between the two halves. 
\subsubsection*{JT Gravity}
The gravitational theory we introduce on $\mathbb H^-$ is Jackiw–Teitelboim (JT) gravity. JT gravity is a two-dimensional theory of gravity that captures universal features of near-extremal black holes and holography in nearly-AdS$_2$ spacetimes \cite{Almheiri:2014cka}. For a nice review of some recent developments, see \cite{Turiaci:2024cad}. Here, will will briefly review JT gravity with a focus on the general solution in Poincar\'e coordinates.

The action of JT gravity is given by
\begin{equation}
     I_\mathrm{JT}[g,\phi]=- S_0 \chi -\frac{1}{16\pi G_\mathrm{N}}\int \mathrm{d}^2\textbf{x}\,\sqrt{g}\,\phi (R+\Lambda) -\frac{1}{8\pi G_\mathrm{N}}\int_{\partial} \mathrm{d}x \,\sqrt{h}\phi(K-1).\label{eq:jt_action}
\end{equation}
with the Euler characteristic \begin{equation}
    \chi=\frac{1}{4\pi}\int\mathrm{d}^2\textbf{x}\,\sqrt{g}R+\frac{1}{2\pi}\int_{\partial} \mathrm{d}x\,\sqrt{h}K
\end{equation}
and $S_0=\frac{1}{4G_\mathrm{N}}\phi_0$. The first term in \cref{eq:jt_action} is called the topological term and can be understood as measuring a ground state entropy.\footnote{Such an interpretation is not always available. In the case of non-supersymmetric near-extremal black holes which are also described by JT gravity coupled to matter, the ground-state degeneracy gets lifted by quantum effects \cite{Iliesiu:2020qvm}.}
The quantity $\Lambda=-2/L^2$ is the cosmological constant and $L$ is the AdS length which we will set to $L=1$ in the following. The one-dimensional integral consists of the Gibbons-Hawking-York boundary term \cite{York:1972sj,Gibbons:1976ue} which needs to be added to make the variational principle well-defined as well as a counter term which is required to obtain a finite on-shell action. While both extra terms are crucial for deriving the Schwarzian boundary dynamics, they are not important for us and we will omit them in the following.

Varying the action \cref{eq:jt_action} with respect to $\phi$ gives the condition \begin{equation}
    R+2=0, \label{R}
\end{equation}
i.e., it constrains the metric to be locally AdS$_2$. A convenient choice of coordinates for AdS$_2$ are Poincar\'e coordinates. In complex notation the metric reads
 \begin{equation}
    \mathrm{d}s^2=-\frac{4}{(z-\overline{z})^2}\mathrm{d}z\mathrm{d}\overline{z}. \label{AdS metric}
\end{equation}
This metric has constant negative curvature $R=-2$, satisfying the JT equations of motion \cref{R}. 

If we perform a variation of $I_\mathrm{JT}[g,\phi]$ with respect to the metric, we obtain the equations of motion for the dilaton\begin{equation}
    \nabla_\mu\nabla_\nu\phi-g_{\mu\nu} \Box \phi+g_{\mu\nu}\phi=0. \label{Hess equation}
\end{equation} 
In the metric \cref{AdS metric} the three equations for our dilaton field read
\begin{align}
    &\left(\frac{\partial}{\partial z}-\frac{\partial}{\partial \overline{z}} \right)^2\phi+\frac{2}{z-\overline{z}}\left(\frac{\partial}{\partial z}-\frac{\partial}{\partial \overline{z}} \right)\phi-\frac{4}{(z-\overline{z})^2}\phi=0, \label{first pon eq} \\ &\left(\frac{\partial^2}{\partial z^2}-\frac{\partial^2}{\partial \overline{z}^2} \right)\phi+\frac{2}{z-\overline{z}}\left(\frac{\partial}{\partial z}+\frac{\partial}{\partial \overline{z}} \right)\phi=0, \\ &\left(\frac{\partial}{\partial z}+\frac{\partial}{\partial \overline{z}} \right)^2\phi+\frac{2}{z-\overline{z}}\left(\frac{\partial}{\partial z}-\frac{\partial}{\partial \overline{z}} \right)\phi+\frac{4}{(z-\overline{z})^2}\phi=0,\label{third pon eq}
\end{align}
with the general solution parametrized by three real constants $\lambda_1, \lambda_2, \lambda_3$,
\begin{equation}
    \phi(z)=2 \mathrm{i} \frac{\lambda_1+\lambda_2\,(z + \bar z)/2 + \lambda_3|z|^2}{z - \bar z} = \frac{\lambda_1+\lambda_2 \tau + \lambda_3(\tau^2 + x^2)}{x}. \label{dilaton}
\end{equation}

In two dimensions, the metric has no degrees of freedom and thus solutions to \cref{R} all locally agree. Therefore, solutions such as global or black hole coordinates, are related to \cref{AdS metric} by (large) diffeomorphisms. 
As will be discussed in \cref{sec:finite_temp}, the metric for a Euclidean black hole and the dilaton solution on this background are related to \cref{AdS metric,dilaton} by the diffeomorphism \eqref{eq:Tan map}, see \cref{eq:ads2_finite_t,eq:dilaton_finite_t}, respectively.

\subsubsection*{Generalized Entropy and Island Entropy}
For a CFT defined on a non-gravitating background, we can define and compute entanglement entropies as explained in the first part of this section. However, in the presence of gravity the natural notion of entropy is not anymore given by the von Neumann entropy of matter fields, $S(A)$, but its correct definition substantially changes in two ways.

First, the natural notion of entropy in the presence of gravity is generalized entropy, which in $d$ dimensions takes the form
\begin{equation}
\label{eq:Sgen}
    S_\mathrm{gen}(A) = \frac{\mathrm{Vol}(\partial A)}{4 G_\mathrm{N}}+S(A),
\end{equation}
where $A$ is a spacelike, co-dimension one subregion and $\mathrm{Vol}(\partial A)$ is the volume of its boundary. Unlike entanglement entropy in a quantum field theory, if $A$ is completely contained in the gravitating part of spacetime generalized entropy is conjectured to be finite \cite{Susskind:1994sm} (for a nice discussion in a more modern parlance, c.f., the appendix of \cite{Bousso:2015mna}). 
In two dimensions, the surface $\partial A$ consists of two points (if $A$ is a single interval) and the volume of its boundary is replaced by the value of the dilaton at the endpoints of $A$.

\begin{figure}[t]
    \centering
\def\svgwidth{15cm}
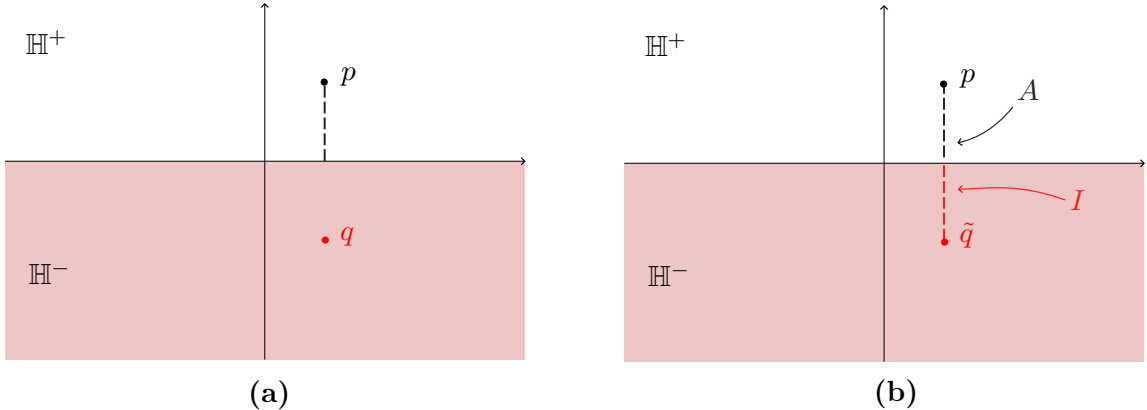
\caption{The upper half-plane is the non-gravitating bath, the lower half-plane is the AdS$_2$ region. \textbf{(a)} The entanglement entropy of an interval $A$, which start at the point $p$ and includes the boundary can be computed by inserting a point $q$ in the AdS$_2$-region and extremizing the generalized entropy of the interval $\tilde{A}=[p,q]$ over $q$. \textbf{(b)} This fixes a point $\tilde{q}$, the quantum extremal surface, and the \emph{island} $I$.
} 
\label{fig:AdS/Bath}
\end{figure}

In the case of an interval in the AdS$_2$/bath system which stretches between the gravitating and non-gravitating region the only contribution to the volume term comes from the endpoint in the gravitating region. Concretely, when we consider an interval $\tilde A = [p,q]$ with endpoints $p\in\mathbb{H}^+$ and $q\in\mathbb{H}^-$, see \cref{fig:AdS/Bath}, we have \begin{equation}
    S_\mathrm{gen}(\tilde A)=\frac{\phi_0+\phi(q)}{4 G_\mathrm{N}}+S_\mathrm{bulk}(\tilde A) \label{generalized}
\end{equation} where $\phi_0$ comes from the topological term in \cref{eq:jt_action}.

Second, the presence of a gravitating subregion does not only modify the definition of the natural notion of entropy for subregions which are within the gravitating subsystem. The correct entropy of a subregion $A$ contained in the bath can receive contributions from additional subregions in the gravitating part of spacetime, so-called islands, which appear dynamically.\footnote{What happens if one specifies a generic subregion in the presence of gravity is not fully understood, but see \cite{Bousso:2022hlz} for a proposal.} The realization that entanglement entropy is modified by the inclusion of additional regions was at the heart of finding the Page-curve of an evaporating black hole from a semi-classical calculation and is nicely encapsulated in the quantum extremal island (QEI) / quantum extremal surface (QES) formula \cite{Engelhardt:2014gca, Almheiri:2019hni,Penington:2019npb}
\begin{align}
    \label{eq:island_entropy}
    S_\mathrm{island}(A) = \underset{I}{\min \operatorname{ext}} \left( \frac{\phi_0+\phi(\partial I)}{4 G_\mathrm{N}} + S_\mathrm{bulk}(A \cup I) \right),
\end{align}
where $A$ is a subregion in the bath and $I$ is a set of codimension one intervals in the gravitating region, see fig \ref{fig:AdS/Bath}. The prescription asks to include any gravitating subregion $I$ in the computation of the generalized entropy if varying this region extremizes the generalized entropy functional and produces an entropy which is smaller than the entropy without including $I$.

\section{Single Interval}
\label{sec:single_interval}
Our goal in this paper is to demonstrate that computing entanglement entropy of subregions in a BCFT with a deformed boundary yields, under appropriate conditions, the same result as computing island entropies of a CFT in an AdS$_2$/bath system, in which the dilaton value is related to the magnitude of the boundary deformation in the BCFT case.

In this section we compute the entropy of a single interval at both zero and finite temperature. We will discuss the zero temperature case first. The case of finite temperature can be obtained from the zero temperature case by a conformal transformation, so the corresponding discussion will be rather brief.

In this section, the intervals we consider in the BCFT description include the boundary (see \cref{fig:interval}). We use the replica trick, reviewed in \cref{sec:review}, to compute the von Neumann entropy as the analytic continuation of a twist operator one-point function in the BCFT. In the AdS$_2$/bath system, we compute the entanglement entropy using the QEI formula, also reviewed in \cref{sec:review}, and find a match between both computations. 

\subsection{Zero Temperature}

\subsubsection{Von Neumann Entropy in BCFTs with Deformed Boundaries}
In order to perform the computation at zero temperature, we start by considering a BCFT$_2$ on the upper half plane $\mathbb{H}^+$.
In the absence of a boundary, the global conformal group of the complex plane $\mathbb{C}$ is $\mathrm{PSL}(2,\mathbb{C})\cong \mathrm{SL}(2,\mathbb C)/\mathbb Z_2$. This group acts on $\mathbb{C}$ via Möbius transformations \begin{equation}
\label{eq:moebius}
    M(z)=\frac{az+b}{cz+d}, \qquad \text{with } 
    \begin{pmatrix}
        a & b \\ c & d
    \end{pmatrix}\in\mathrm{PSL}(2,\mathbb{C}). 
\end{equation} 
Restricting to those transformations compatible with the BCFT$_2$ boundary, i.e., those transformations which map the upper half-plane to itself, breaks $\mathrm{PSL}(2,\mathbb{C})$ to the $\mathrm{PSL}(2,\mathbb{R})$ subgroup which maps the real line to itself, i.e., \cref{eq:moebius} with all coefficients real. The broken transformations are elements of $\mathrm{PSL}(2,\mathbb{C})/\mathrm{PSL}(2,\mathbb R)$ and are not symmetries of the BCFT anymore, but instead relate BCFTs with different boundary shapes. To obtain a background with a boundary deformed by a broken global conformal transformation, we can apply a Möbius transformation for which not all coefficients are real to the upper half-plane.

Since we are interested in entanglement entropy of BCFTs whose boundaries have been deformed at the linear level, it is sufficient for us to look at infinitesimal global conformal transformations. They take the form
\begin{align}
\label{eq:conf_trans}
    z \mapsto z + \xi(z), \hspace{0.5cm} \text{with} \hspace{0.5cm} \xi(z) = \mathrm i \tilde{\lambda}_1 + \mathrm i \tilde{\lambda}_2 z  + \mathrm i \tilde{\lambda}_3 z^2
\end{align} with $\tilde{\lambda}_1,\tilde{\lambda}_2,\tilde{\lambda}_3\in\mathbb{C}$ which can be obtained by expanding \cref{eq:moebius} around the identity. 
Again, only those infinitesimal transformations for which the coefficients $\mathrm{i} \lambda_i$ are all real (i.e., $\lambda_i$ are imaginary) leave the real line invariant and are a symmetry of the BCFT. In practice, to apply a deformation of the boundary it is simplest to choose $\tilde{\lambda}_1,\tilde{\lambda_2},\tilde{\lambda}_3$ all real, which ensures that the deformation is orthogonal to the boundary. This convention is the reason for the factors of $\mathrm{i}$ in \cref{eq:conf_trans} which simplifies equations below slightly.

\begin{figure}[t]
    \centering
\def\svgwidth{11cm}
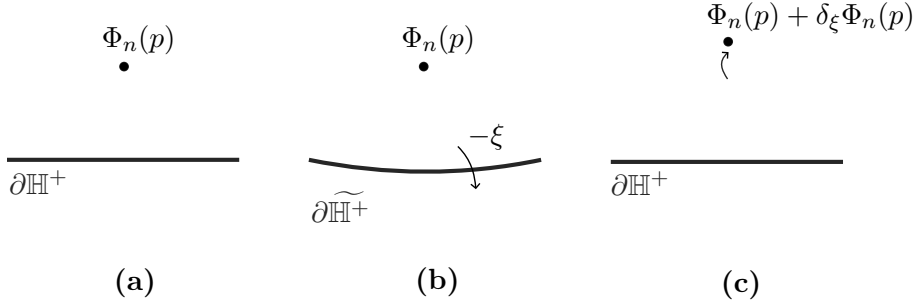
\caption{\textbf{(a)} To compute R\'enyi entropies, we need the expectation value of a twist operator in the upper half-plane. The effect of a deformation of the boundary \textbf{(b)} on the twist one-point function can also be captured by displacing the twist operator in the BCFT with an undeformed boundary \textbf{(c)}.
}
\label{fig:boundary_deformation}
\end{figure}

We now deform the BCFT boundary by $-\xi$. In the deformed geometry $\widetilde {\mathbb H^+}$, obtained by acting with \cref{eq:conf_trans} on $\mathbb H^+$, we now select a single interval $A$ by choosing a point $p$ and connecting it to the boundary. We evaluate the von Neumann entropy on the deformed geometry, which we denote by $\tilde S(A)$, by inserting a twist operator $\Phi_n(p)$ in $\widetilde {\mathbb H^+}$ at $p$, and computing and analytically continuing the R\'enyi entropy as explained in \cref{sec:review}. For the R\'enyi entropy we need the twist one-point function in the presence of a deformed boundary. To evaluate it we act with the inverse of \cref{eq:conf_trans}, mapping the background geometry $\widetilde {\mathbb H^+}$ back to the upper half-plane with a planar boundary, ${\mathbb H^+}$, see \cref{fig:boundary_deformation}. This will also act on $\Phi_n(p)$ and the correction to the R\'enyi entropy is given in terms of the correction to the correlator $\braket{\delta_{\xi}\Phi_n(p)}_{\mathbb H^+}$, i.e.,
\begin{align}
    \tilde S^{(n)}(A) =  S^{(n)}(A) + \frac 1 {1-n}  \log \braket{\delta_{\xi} \Phi_n(p)}_{\mathbb H^+},
\end{align}
where $\delta_\xi$ is the change of $\Phi_n$ under an infinitesimal transformation.

We can arrive at the same conclusion in a slightly different way which is more insightful in the current context. As explained in the introduction, holographic considerations suggest that the boundary value of the dilaton sources the displacement operator \cite{Billo:2016cpy,Miao:2018dvm}. The displacement operator is the response of the system to deforming the boundary, i.e.
\begin{align}
    \delta_X \braket{\dots }_{\mathbb H^+} = \frac 1 {2 \pi} \int \mathrm{d}\tau \Delta X(\tau) \braket {\mathcal D(\tau) \dots}_{\mathbb H^+} ,
    \label{rrr}
\end{align}
where $\delta_X$ denotes a change in the embedding of the boundary by $\Delta X(\tau)$ into the lower half plane, i.e., into the direction of the outward pointing normal $- \partial_x$. Its precise form can be derived from the local form of the conformal Ward identities in the presence of a boundary. Doing so yields the simple relation 
\begin{align}
    \nabla_\mu T^{\mu n} =  - \delta(x) \mathcal  D(\tau),\label{displdef}
\end{align}
where $n$ denotes the outwards oriented normal. Integrating this equation show that the displacement operator is the normal-normal component of the stress-energy tensor at the boundary. In our BCFT$_2$ setup, this means: \begin{equation}
    \mathcal{D}(\tau)=(T+\overline{T})(\tau).
\end{equation} 
We can deform the boundary in $-x$ direction by an amount $\Delta X(\tau)$, where
\begin{align}
\label{eq:def_trans}
\Delta X(\tau) = -\mathrm{i} \xi(\tau)
\end{align}
by inserting\footnote{Our expressions differ from expressions in \cite{Billo:2016cpy} by a factor of $(2 \pi)^{-1}$. This is because we follow the usual CFT$_2$ conventions which include a factor of $2\pi$ in the definition of the stress-energy tensor and consequently also the displacement operator.}
\begin{align}
    \label{eq:boundary_deformation}
    \exp\left(\frac{1}{2\pi \mathrm{i}}\int \mathrm{d}\tau\,\xi(\tau) \mathcal D(\tau)\right)
\end{align}
into correlation functions. Here and in the following, we assume that the coefficients of $\xi$ are imaginary such that the displacement is real. Since we are only interested in deformations at linear order, it is sufficient for us to use the expansion of \cref{eq:boundary_deformation} up to first order.

The conclusion is of course unchanged with respect to the previous argument. We can use the fact that the displacement operator is the normal-normal component of the stress-energy tensor to show that 
\begin{align}
\label{eq:displacement_ward} 
\frac{1}{2\pi \mathrm{i}} \int \mathrm{d}\tau\, \xi(\tau) \braket{\mathcal D(\tau) \prod_{j=1}^n\mathcal{O}_j(z_j,\bar{z}_j)}  =  \braket{\delta_\xi \prod_{j=1}^n\mathcal{O}_j(z_j,\bar{z}_j)}.
\end{align}

Therefore, following either argument, the correction to the twist one-point function due to a deformation of the boundary is given by
\begin{equation}
    \braket{\Phi_n(p,\bar{p})}_{\widetilde{\mathbb{H}^+}} = \braket{\Phi_n(p,\bar{p})}_{\mathbb{H}^+} + \braket{\delta_{\xi}\Phi_n(p,\bar{p})}_{\mathbb{H}^+}, 
\end{equation}
where $\xi$ is a broken global conformal generator which undoes the boundary deformation.

The transformation behavior of a primary operator such as the twist is
\begin{equation}
        \braket{\delta_{\xi}\Phi_n(p,\bar{p})}_{\mathbb{H}^+} = \left(h\xi'(p)+\xi(p)\partial_{p}-h\xi'(\bar{p})-\xi(\bar{p})\partial_{\bar{p}}\right)\braket{\Phi_n(p,\bar{p})}_{\mathbb{H}^+}. \label{infTrafoCorr}
    \end{equation}
Using that the coefficients $\tilde \lambda_i$ of $\xi$ are chosen real, this can be rewritten as
\begin{equation}
\label{eq:change_one_pt_function}
\braket{\delta_{\xi}\Phi_n(p,\bar{p})}_{\mathbb{H}^+}=d_n\left(\mathrm{Re}\,\xi'(p) - \frac{1}{\mathrm{Im}\,p}\mathrm{Im}\,\xi(p) \right)\braket{\Phi_n(p,\bar{p})}_{\mathbb{H}^+}.
\end{equation}
From this, the linear correction to the $n$-th R\'enyi entropy follows 
and the analytic continuation $\lim_{n\to 1} \delta S^{(n)}$ produces the first correction to entanglement entropy \begin{equation}
    \delta \tilde S(A) = - \frac{c}{6}\left(\mathrm{Re}\,\xi'(p) - \frac{1}{\mathrm{Im}\,p}\mathrm{Im}\,\xi(p) \right).
\end{equation}
We can evaluate the correction and find
\begin{equation}
    \mathrm{Re}\,\xi'(p) - \frac{1}{\mathrm{Im}\,p}\mathrm{Im}\,\xi(p) = - \frac{\tilde{\lambda}_1+\tilde{\lambda}_2\mathrm{Re}\,p+\tilde{\lambda}_3|p|^2}{\mathrm{Im}\,p}.
\end{equation}
The full expression for the entanglement entropy for an interval in the presence of a deformed boundary then reads (up to first order) \begin{equation}
    \tilde S(A) = \frac{c}{6}\log\frac{2p_x}{\epsilon}+\log g_b+\frac{c}{6}\frac{\tilde{\lambda}_1+\tilde{\lambda}_2 p_\tau+\tilde{\lambda}_3(p_\tau^2+p_x^2)}{p_x}. \label{defo}
\end{equation}
As we can see by comparing to \cref{eq:def_trans}, the coefficients of the correction are exactly set by the deformation of the boundary.

\subsubsection{Island Entropy of the AdS/Bath-System}
\label{subsubsec: Island entropy of AdS bath system}
We now want to compare the expression \cref{defo} to an island entropy computation in a AdS$_2$/bath system, where the AdS$_2$ region is coupled to JT gravity, c.f.\ \cref{Subsec AdS$_2$/Bath Geometry and generalized Entropy}. We choose the upper half-plane, $x\geq 0$, to be our non-gravitating bath, i.e., it is equipped with the ordinary Euclidean metric, which we denote by $h$, 
\begin{equation}
    \mathrm{d}s_h^2=\mathrm{d}z\mathrm{d}\overline{z}.
\end{equation}
The lower half-plane $x<0$ is equipped with the AdS$_2$-metric $g$, \begin{equation}
    \mathrm{d}s_g^2=-\frac{4}{(z-\overline{z})^2}\mathrm{d}z\mathrm{d}\overline{z},
\end{equation} 
and coupled to JT gravity, \cref{fig:AdS/Bath}. Note that unlike in the BCFT computation, the boundary of the non-gravitating region is not deformed, but simply the real line.

In the AdS$_2$/bath description, we consider the same interval endpoint $p$ as in the BCFT description of our system. Because of the absence of a boundary we have a second interval endpoint in the lower half-plane, which we call $q$. Via the island entropy formula \cref{eq:island_entropy}, the second endpoint is dynamically determined and lies in the gravitating region.

The computation of the island entropy follows the by now well-known prescriptions developed in \cite{Almheiri:2019psf} and reviewed in \cref{sec:review} and proceeds in two steps. First we compute the generalized entropy for an arbitrary endpoint $q$. Then, in a second step we extremize over this second endpoint in the gravitating region.\footnote{Of course, it could be that one also needs to consider additional, disjoint interval configuration. This is however not the case here.}

To compute the generalized entropy, \cref{generalized}, we first need the bulk entanglement entropy $S_\mathrm{bulk}([p,q])$ which we can obtain from a CFT twist two-point function with $\Phi_n$ and $\overline{\Phi}_n$ inserted at $q$ and $p$, respectively. Specializing the general form of a primary two-point function on the complex plane
\begin{equation}
\braket{\mathcal{O}_1(z_1)\mathcal{O}_2(z_2)}_\mathbb{C}=\frac{\mathscr{C}_{\mathcal{O}_1\mathcal{O}_2}}{|z_1-z_2|^{2\Delta}},
\end{equation} 
to the case of twists at $p = z_1$ and $q = z_2$ with
\begin{equation}
    \label{eq:norm_twist_two_point}
    \mathscr{C}_{\overline{\Phi}_n\Phi_n}=(\braket{a|0}\braket{0|a})^{(1-n)}\epsilon^{2d_n}.
\end{equation}
we obtain \begin{equation}
\label{eq:twist_two_pt_plane}
    \braket{\overline{\Phi}_n(q)\Phi_n(p)}_\mathbb{C}=(\braket{a|0}\braket{0|a})^{(1-n)}\left(\frac{|q-p|}{\epsilon}\right)^{-2d_n}.
\end{equation} 
\Cref{eq:norm_twist_two_point} is obtained by cutting out small disks around $p$ and $q$ and mapping the interval $[p, q]$ to a cylinder, analogous to the discussion in \cref{subsec Ent}. The boundary state $\ket a$ captures the boundary conditions at the disk boundaries and can be absorbed in the regulator $\epsilon$.

To arrive at the correlator in the AdS$_2$/bath system, we need to take into account that part of the interval $[p,q]$ lives in the AdS region for which the metric differs from the Euclidean metric by a Weyl transformation. Under a Weyl transformation
\begin{equation}
    g=\Omega^2 h
\end{equation} 
an $n$-point correlation function transforms as 
\begin{equation}
    G^g(z_1, \dots, z_n)=|\Omega(z_1)|^{-\Delta}\dots|\Omega(z_n)|^{-\Delta}G^h(z_1, \dots, z_n).
\end{equation}
In our case, only the lower half-plane is Weyl transformed and thus only the point $q$ is affected.

Having a point $p\in\mathbb{H}^+$ and a point $q\in \mathbb{H}^-$, we can compute the bulk correlation function between the two points $p$ and $q$, given by
\begin{equation}
    \braket{\overline{\Phi}_n(q)\Phi_n(p)}_\mathrm{AdS_2/bath}=(- q_x )^{d_n}\;\frac{\mathscr{C}_{\overline{\Phi}_n\Phi_n}}{|p-q|^{2d_n}},
\end{equation} since the metric $g$ in \cref{AdS metric} is related to the Euclidean metric $\eta$ by $\Omega(z)^2=\frac{1}{x^2}$. The minus sign accounts for the fact that $q$ lives in the lower half-plane and thus $q_x = \mathrm{Im} \;q<0$.

The R\'enyi entropy of the interval $[p,q]$ is then given by \begin{equation}
    S_\mathrm{bulk}^{(n)}([p,q])=\frac{1}{1-n}\log\mathscr{C}_{\overline{\Phi}_n\Phi_n}+\frac{c}{12}\frac{n+1}{n}\log\frac{|p-q|^2}{-q_x}
\end{equation}
and the bulk entanglement entropy reads

\begin{equation}
    \label{eq:bulk_entropy}
    S_\mathrm{bulk}([p,q])=\frac{c}{6}\log\frac{|p-q|^2}{-\epsilon^2 q_x},
\end{equation} 
where we have used \cref{eq:norm_twist_two_point} and redefined the cutoff to absorb the boundary state overlaps, i.e.,
\begin{equation}
    \lim_{n\to 1}\frac{1}{1-n}\log\mathscr{C}_{\overline{\Phi}_n\Phi_n}=-\frac{c}{6}\log \epsilon^2. \label{CUT}
\end{equation}

Since the lower half-plane hosts JT gravity, the generalized entropy is given by \cref{eq:bulk_entropy}, together with the dilaton evaluated at point $q$, i.e.,
\begin{equation}
    \label{eq:bulk_gen_entropy}
    S_\mathrm{gen}([p,q])=\frac{1}{4G_\mathrm{N}}(\phi_0+\phi(q))+S_\mathrm{bulk}([p,q]),
\end{equation} where $\phi_0$ is a constant. The non-constant contribution $\phi(q)$ is a general solution to the dilaton equation of motion for JT on the Poincar\'e patch, given in \cref{dilaton},
\begin{equation}
    \phi(q)=\frac{\lambda_1+\lambda_2\,\mathrm{Re}\,q+ \lambda_3|q|^2}{-\mathrm{Im}\,q}=\frac{\lambda_1+\lambda_2\,q_\tau + \lambda_3(q_\tau^2+q_x^2)}{-q_x}.
\end{equation}
The island entropy is then the smallest extremum of \cref{eq:bulk_gen_entropy},
\begin{equation}
    S_\mathrm{island}(p)=\underset{q\in\mathbb{H}^-}{\min \operatorname{ext}} \; S_\mathrm{gen}([p,q]). \label{Island Formula}
\end{equation}

Instead of directly evaluating the island entropy in the most general case, it is useful to start with the case for which the non-trivial part of the dilaton is turned off.
Here, the generalized entropy simply takes the form \begin{equation}
    S_\mathrm{gen}([p,q])=\frac{1}{4G_\mathrm{N}}\phi_0+\frac{c}{6}\log\frac{|p-q|^2}{-\epsilon^2\,q_x} \label{trivial dil S}
\end{equation} 
and is extremized with respect to $q$ for
\begin{equation}
    \label{eq:extremum}
    \tilde q =\bar{p}.
\end{equation}
Replacing $\tilde q \to \bar{p}$ in \eqref{trivial dil S}, the island entropy becomes
\begin{equation}
    S_\mathrm{island}(p)=\frac{1}{4G_\mathrm{N}}\phi_0+\frac{c}{6}\log\frac{4 p_x}{\epsilon^2}
\end{equation}

Finally, we can consider a general dilaton profile. In this case, we would have to extremize the generalized entropy with arbitrary coefficients $\lambda_i$. However, at linear order in $\lambda_i$ the correct value of the island entropy is simply given by evaluating the generalized entropy, \cref{eq:bulk_gen_entropy}, including a non-trivial dilaton profile at the old extremum, \cref{eq:extremum}.

The reason is that since the expression for the generalized entropy without a non-trivial dilaton profile, \cref{trivial dil S}, is extremal for $\tilde q = \bar{p}$, linear variations of the location of the quantum extremal surface do no change its value. Moreover, the additional term coming from the non-trivial dilaton solution is already linear in $\lambda_i$ and therefore it is sufficient to evaluate it at the zeroth order extremum. 
This leaves us with the final form of the island entropy
\begin{align}
    \label{eq:island_entropy_plane}
     S_\mathrm{island}(p)=\frac{c}{6}\log\frac{2 p_x}{\epsilon} + \frac{1}{4G_\mathrm{N}}\left(\phi_0+\frac{\lambda_1+\lambda_2\,p_\tau + \lambda_3(p_\tau^2+p_x^2)}{p_x}\right)+\frac{c}{6}\log\frac{2}{\epsilon}.
\end{align}

\subsubsection{Matching BCFT and AdS/Bath Entropies}
We can now compare the results of the BCFT and AdS$_2$/bath computations. As explained in \cref{sec:intro}, we expect both expressions to agree if we choose the gravitational constant according to \cref{eq:gnewton} and the displacement to be set by the asymptotic value of the dilaton, \cref{eq:deformation_expectation}.
And in fact, it can easily be checked that the linear corrections in \cref{defo} and \cref{eq:island_entropy_plane} agree if the dilaton solution is set by the deformation, i.e., 
\begin{align}
    \label{eq:dilaton_parameter_matching_plane}
    \lambda_i = \tilde \lambda_i = \Delta X_i,
\end{align}
where we have denoted the corresponding coefficient in the deformation \eqref{eq:def_trans} by $\Delta X_i$.
The identification \cref{eq:dilaton_parameter_matching_plane} precisely agrees with our expectation, \cref{eq:deformation_expectation}. 

Matching the leading order terms we find a relation between the boundary entropy on the BCFT side and the dilaton in the AdS$_2$/bath system,
\begin{align}
    \log g_b = \frac{1}{4G_\mathrm{N}}\phi_0+\frac{c}{6}\log\frac{2}{\epsilon}. \label{Bound entrop} 
\end{align}
Naively, the appearance of a cutoff on the right-hand side of \cref{Bound entrop} might be worrying, since the left-hand side is finite. However, recall that $\phi_0/4 G_\mathrm{N}$ is the bare quantity that appears in the Lagrangian and obtains corrections from matter fields, which are encapsulated in the second, $c$-dependent term. Thus the right hand side of \cref{Bound entrop} should really be thought of as the renormalized coefficient of the topological term in JT.

In the construction of \cite{Neuenfeld:2024gta} which motivated our computation, this can be explicitly seen. There, a JT Lagrangian on a two-dimensional end-of-the-world brane was derived by integrating out the AdS$_3$ bulk. The leading order coefficient of the Euler characteristic is shifted due to quantum corrections from matter fields by a contribution proportional to $\log \epsilon$, where $\epsilon$ plays a role of the UV cutoff on the brane. Thus, \cref{Bound entrop} is the statement that the boundary entropy equals the effective, renormalized dilaton including quantum corrections. The result \eqref{Bound entrop} should have been expected, since both sides measure a ground state degeneracy in the respective theory.



\subsection{Finite Temperature}

\label{sec:finite_temp}
\subsubsection{Von Neumann Entropy in BCFTs with Deformed Boundaries}
The result obtained above can readily be translated to the case of finite temperature by considering a BCFT on a cylinder of circumference $\beta$. The most straight-forward way of doing this is by choosing coordinates $w, \bar w$ which are related to coordinates $z, \bar z$ on the upper half-plane through the conformal transformation
\begin{equation}
    z=\tan\frac{\pi w}{\beta},\label{eq:Tan map}
\end{equation}
and similarly for $\bar z$ and $\bar w$. The new coordinates take values in the range $\mathrm{Im}\,w \geq 0$ and $\mathrm{Re}\,w \in (-\beta/2, \beta/2]$ is periodically identified with period $\beta$.

The infinitesimal broken conformal transformations now read
\begin{align}
    \xi^\beta(w) = \frac {\mathrm{i}\beta} {2 \pi} \left(\tilde \mu_1 + \tilde \mu_2 \sin\left(\frac{2 \pi w}{\beta}\right) + \tilde \mu_3 \cos\left(\frac{2 \pi w}{\beta}\right) \right),
\end{align}
where $\tilde \mu_i$ are related to $\tilde \lambda_i$ as
\begin{align}
\label{eq:mu_lambda}
\tilde \mu_1  = \tilde \lambda_1 + \tilde \lambda_3,\qquad \tilde \mu_2  =  \tilde \lambda_2, \qquad
\tilde \mu_3  = \tilde \lambda_1 - \tilde \lambda_3.
\end{align}
As can be seen from these expressions, transformations which move the boundary in a normal direction again correspond to purely real coefficients.

The correction to the entanglement entropy of an interval which extends from a single point $p$ can then simply be obtained from the transformation behavior of the twist one-point function,
  \begin{equation}
   \braket{\Phi_n(p,\bar{p})}_{\mathbb{H}^+_\beta}=\mathscr{A}_{\Phi_n}\left(\frac{\beta}{\pi}\sinh\frac{2\pi\,p_x}{\beta}\right)^{-d_n}. \label{corrTemp}
\end{equation} 
Translating this to the R\'enyi entropy and taking the $n \to 1$ limit produces the von Neumann entropy for an interval $A$ from $p$ perpendicular to the boundary \begin{equation}
    S^{\beta}(A)=\frac{c}{6}\log\left(\frac{\beta}{\pi\epsilon}\sinh\frac{2\pi p_x}{\beta}\right)+\log g_b.
\end{equation}

Then, to obtain corrections which arise from boundary deformations we compute again the infinitesimal transformation of the correlation function $\braket{\Phi_n(p,\bar{p})}_{\mathbb{H}^+_\beta}$ under broken conformal transformations. Since the boundary is the same as before, namely the real axis (only equipped with a $\beta$-periodicity), the computation is completely analogous to the one above. They key element is the correction to the twist one-point function \begin{equation}
    \braket{\delta_{\xi^\beta}\Phi_n(p,\overline{p})}_{\mathbb{H}^+_\beta}=d_n\left(\mathrm{Re}\,\xi^\beta\,'(p) -\frac{2\pi}{\beta}\mathrm{coth}\frac{2\pi p_x}{\beta}\,\mathrm{Im}\,\xi(p)\right)\braket{\Phi_n(p,\overline{p})}_{\mathbb{H}^+_\beta}.
\end{equation}
From this, we can compute the correction to the R\'enyi and eventually von Neumann entropy. Using the explicit form of $\xi^\beta(w)$ we find the following result for the entanglement entropy to first order in the boundary deformation: 
\begin{align}
\begin{split}
    \tilde{S}^\beta(A)=&\frac{c}{6}\log\left(\frac{\beta}{\pi\epsilon}\sinh\frac{2\pi p_x}{\beta}\right)+\log g_b \\ &+ \frac{c}{6}\frac{1}{\sinh\frac{2\pi p_x}{\beta}}\left(\tilde{\mu}_1\cosh\frac{2\pi p_x}{\beta}+\tilde{\mu}_2\sin\frac{2\pi p_\tau}{\beta}+\tilde{\mu}_3\cos\frac{2\pi p_\tau}{\beta}\right).
    \label{FiniteTemperatureEntPer}
    \end{split}
\end{align}

\subsubsection{Island Entropy of the AdS/Bath-System}
As we did for the zero temperature case, we want to link equation \eqref{FiniteTemperatureEntPer} to the island entropy in the AdS$_2$/bath system, where we again want to attach a dilaton in the AdS$_2$ region. Again as in the zero temperature case, we have the non-gravitating bath at $x\geq 0$, equipped with the Euclidean metric \begin{equation}
    \mathrm{d}s^2=\mathrm{d}w\mathrm{d}\overline{w}.
\end{equation}
The AdS$_2$-region at finite temperature can be obtained from the flat case using the diffeomorphism \eqref{eq:Tan map} and reads
\begin{equation}
    \label{eq:ads2_finite_t}
    \mathrm{d}s^2_\beta=-\frac{4\pi^2}{\beta^2}\frac{1}{\sin^2\frac{\pi(w-\overline{w})}{\beta}}\mathrm{d}w\mathrm{d}\overline{w},
\end{equation}
which is defined for $\mathrm{Im}\,w<0$.

We again fix a point $p$ in the bath and consider a point $q$ in the AdS$_2$-region. Using the island formula we then determine the location $\tilde q$ which extremizes the generalized entropy of the interval $[p,q]$. For the generalized entropy with have to compute the bulk entanglement entropy $S_\mathrm{bulk}^\beta([p,q])$, obtained from a twist and anti-twist two point function. For the finite temperature case, we have to transform $\braket{\overline{\Phi}_n(q)\Phi_n(p)}_{\mathrm{AdS}_2/\mathrm{bath}}$ under the map \eqref{eq:Tan map},
\begin{equation}
\braket{\overline{\Phi}_n(q)\Phi_n(p)}_{\mathrm{AdS}_2/\mathrm{bath}}^\beta=\mathscr{C}_{\overline{\Phi}_n\Phi_n}\left(\frac{\pi}{2\beta}\frac{-\sinh\frac{2\pi q_x}{\beta}}{\sin^2\frac{\pi(p_\tau-q_\tau)}{\beta}+\sinh^2\frac{\pi(p_x-q_x)}{\beta}}\right)^{d_n},
\end{equation}
from which we obtain the bulk R\'enyi entropy
 \begin{equation}
    S^{(n)\beta}_\mathrm{bulk}([p,q])=\frac{1}{1-n}\log\mathscr{C}_{\overline{\Phi}_n\Phi_n}+\frac{c}{12}\frac{n+1}{n}\log\left(\frac{2\beta}{\pi}\frac{\sin^2\frac{\pi(p_\tau-q_\tau)}{\beta}+\sinh^2\frac{\pi(p_x-q_x)}{\beta}}{-\sinh\frac{2\pi q_x}{\beta}}\right),
\end{equation} 
as well as the entanglement entropy \begin{equation}
    S^{\beta}_\mathrm{bulk}([p,q])=\frac{c}{6}\log\left(\frac{2\beta}{\pi\epsilon^2}\frac{\sin^2\frac{\pi(p_\tau-q_\tau)}{\beta}+\sinh^2\frac{\pi(p_x-q_x)}{\beta}}{-\sinh\frac{2\pi q_x}{\beta}}\right).
\end{equation} 
The boundary state overlaps in \cref{eq:norm_twist_two_point} are again absorbed by redefinition of the cut off.
With respect to JT gravity in the lower half plane, the generalized entropy for our system comes again with a dilaton contribution at the point $q$, and is given by \begin{equation}
    S^\beta_\mathrm{gen}([p,q])=\frac{1}{4 G_\mathrm{N}}\left(\phi_0+\phi^\beta(q)\right)+S^\beta_\mathrm{bulk}([p,q]). \label{gene2}
\end{equation}
It can be obtained from the dilaton solution at zero temperature \cref{dilaton} by applying the coordinate transformation \cref{eq:Tan map} and using \cref{eq:mu_lambda}. The general form can of course also be obtained by solving the dilaton equations of motion \cref{Hess equation} in the metric \cref{eq:ads2_finite_t}. It reads
\begin{align}
    \label{eq:dilaton_finite_t}
    \phi^\beta(q)
    &=\frac{\mu_1\cosh\frac{2\pi q_x}{\beta}+\mu_2\sin\frac{2\pi q_\tau}{\beta}+\mu_3\cos\frac{2\pi q_\tau}{\beta}}{-\sinh\frac{2\pi q_x}{\beta}}.
\end{align}
Extremization via 
\begin{equation}
    S^\beta_\mathrm{island}(p)=\underset{q\in\mathbb{H}^-}{\min \operatorname{ext}} \; S^\beta_\mathrm{gen}([p,q]).
\end{equation}
again gives the island entropy. In the case of a vanishing dilaton field, the generalized entropy \begin{equation}
    S^\beta_\mathrm{gen}([p,q])=\frac{1}{4 G_\mathrm{N}}\phi_0+\frac{c}{6}\log\left(\frac{2\beta}{\pi\epsilon^2}\frac{\sin^2\frac{\pi(p_\tau-q_\tau)}{\beta}+\sinh^2\frac{\pi(p_x-q_x)}{\beta}}{-\sinh\frac{2\pi q_x}{\beta}}\right)
\end{equation} is extremized at \begin{equation}
    \tilde{q}=\overline{p}. \label{ext2}
\end{equation} 
Completely analogous to the zero temperature case, we can use this information to evaluate the island entropy in the presence of a non-trivial dilaton profile at leading order
\begin{align}
    \label{eq:finite_t_result_ads}
    S^\beta_\mathrm{island}(p) =&\frac{1}{4 G_\mathrm{N}}\left(\phi_0+\frac{\mu_1\cosh\frac{2\pi p_x}{\beta}+\mu_2\cos\frac{2\pi p_\tau}{\beta}+\mu_3\sin\frac{2\pi p_\tau}{\beta}}{\sinh\frac{2\pi p_x}{\beta}}\right)\notag \\ +&\frac{c}{6}\log\frac{2}{\epsilon}+
    \frac{c}{6}\log\left(\frac{\beta}{\pi\epsilon}\sinh\frac{2\pi p_x}{\beta} \right).
\end{align}

\subsubsection{Matching BCFT and AdS/Bath Entropies}
The matching between the BCFT and the AdS$_2$/bath results, now for finite temperature, works as before. We again find that 
\begin{equation}
     \log g_b = \frac{1}{4G_\mathrm{N}}\phi_0+\frac{c}{6}\log\frac{2}{\epsilon},\label{Bound ent temp}
\end{equation}
and using $G_\mathrm{N}=\frac{3}{2\mathrm{c}}$ we conclude that the \cref{FiniteTemperatureEntPer,eq:finite_t_result_ads} agree if
\begin{equation}
    \mu_i=\tilde{\mu}_i
\end{equation} 
which agrees with our expectation that we can identify the asymptotic value of the dilaton with the boundary deformation, i.e., equation \eqref{eq:deformation_expectation} holds.

\section{Two Intervals}
\label{sec:multiple_intervals}
In our previous discussion we restricted ourselves to BCFT one-point functions and CFT two-point functions, respectively. In that case no details of the operator spectrum of the underlying CFT entered the discussion. The reason is that all correlation functions involved are, up to an overall scaling, completely determined by symmetry \cite{Calabrese:2004eu,Calabrese:2009qy}. To test our proposal beyond this simple limit and identify necessary conditions such that a description of boundary deformations in terms of JT gravity emerges, we now want to consider a situation in which details of the underlying CFT cannot be ignored. This is the case for two- and four-point functions of twist operators in the BCFT and CFT, respectively, where sums over conformal blocks enter \cite{Calabrese:2009ez, Calabrese:2010he}.

Since we want the von Neumann entropy in a boundary-deformed BCFT to be reproduced by the island formula, we in particular require that the resulting expression reduces to a sum over independent contributions from intervals. It was shown in \cite{Hartman:2013mia,Sully:2020pza} that for a CFT and BCFT, respectively, this is universally the case at large $c$. Moreover, our approach implicitly assumes that we can treat JT gravity as a semi-classical theory, which implies that the coupling must be small, or equivalently, the central charge must be large, c.f.\ \cref{eq:gnewton}. From here on we will therefore assume large $c$. In \cite{Hartman:2013mia,Sully:2020pza} the authors additionally required that the entropies have a holographic interpretation, i.e., can be computed by RT surfaces \cite{Ryu:2006bv} in a higher-dimensional bulk. This put additional constraints on the spectrum which do not need to apply in our case.

For our computation we place two twist operators at the points, $p, p' \in \mathbb H^+$ and introduce the boundary conformal cross-ratio 
\begin{equation}
    \label{eq:bcft_cross_ratio}
    \eta=\frac{(p-p^*)(p'-{p'}^*)}{(p-{p'}^*)(p'-p^*)},
\end{equation} 
which is invariant under conformal transformations that preserve the location of the boundary. As can be easily seen from this expression, $\eta \to 0$ corresponds to the limit in which the distance between the points $p, p'$ is much bigger then their distance to the boundary, while $\eta \to 1$ corresponds to the opposite limit. 

The twist BCFT two-point function can be expanded in conformal blocks, see \cref{sec:BCFT}. In the limit $\eta \to 1$ it is natural to expand in the bulk channel which, under appropriate conditions and in a sense to be described below, in the large-$c$ limit is completely controlled by the vacuum block. As $\eta \to 0$ similar statements hold for the boundary channel. As shown in \cite{Sully:2020pza}, in the large-$c$ limit, the dominance of the vacuum block in the bulk channel translates holographically to a situation where no island phase is present while in the boundary channel it corresponds to an island contribution in the holographic computation.

We will therefore organize our computation in BCFT channels. First we compute the bulk channel in the presence of a deformed boundary and ask which conditions we have to impose to find a result which is consistent with a AdS$_2$/bath computation without an island contribution. After this, we will repeat the computation in the boundary channel and compute the correction due to a deformed boundary which can be matched to the JT result.

\subsection{Bulk Channel}

Let us start by considering the BCFT bulk channel. We place two twist operators at the points $p, p' \in \mathbb H^+$ and expand the two-point function by first using the bulk OPE and then the BOE. The result can be expressed as a sum over bulk primaries with dimensions $\Delta_i$ involving conformal blocks $\mathscr{F}$, 
\begin{align}
\label{eq:bcft_bulk_expansion}
    \braket{\Phi_n(p,\bar{p})\overline{\Phi}_n(p',\bar{p}')}_{\mathbb{H}^+} =\left(\frac{1-\eta}{|p-p'|^2}\right)^{d_n}\sum_i\mathscr{C}_{\Phi_n\overline{\Phi}_n}^{i}\mathscr{A}_i\mathscr{F}(c,\Delta_i,d_n/2\,|\,1-\eta),
\end{align} 
where we used \eqref{crossRel} with the conformal cross ratio \eqref{eq:bcft_cross_ratio}. The coefficients $\mathscr{C}_{\Phi_n\overline{\Phi}_n}^{i}$ are the bulk CFT OPE coefficients and as above $\mathscr{A}_i$ are one-point function coefficients in the presence of a boundary.

In order to agree with a JT gravity island entropy computation where no island appears we need that the entanglement entropy does not see the effect of the boundary deformation. In addition, it needs to be compatible with the CFT entanglement entropy of the interval $A=[p,p']$ when considered on the whole complex plane. We will now implement these conditions.

The response of the twist correlator to an infinitesimal boundary deformation is given by 
\begin{align}
\label{eq:var_bcft_two_pt}
{}&\braket{\delta_{\xi}\left(\Phi_n(p,\bar{p})\overline{\Phi}_n(p',\bar{p}')\right)}_{\mathbb{H}^+}=\left( \frac{d_n} 2  \xi'(p)+\xi(p)\partial_{p}- \frac{d_n} 2 \xi'(\bar{p})-\xi(\bar{p})\partial_{\bar{p}} \right.\notag \\ &\qquad +\left. \frac{d_n} 2 \xi'(p')+\xi(p')\partial_{p'}-\frac{d_n} 2 \xi'(\bar{p}')-\xi(\bar{p}')\partial_{\bar{p}'}\right)\braket{\Phi_n(p,\bar{p})\overline{\Phi}_n(p',\bar{p}')}_{\mathbb{H}^+}.
\end{align}
It is convenient to introduce the following notation
\begin{equation}
        K_{\mathrm{blk}}(\eta)=\sum_i\mathscr{C}_{\Phi_n\overline{\Phi}_n}^{i}\mathscr{A}_i\,\mathscr{F}(c,\Delta_i,d_n/2\,|\,1-\eta),
    \end{equation}
which captures the conformal block contribution of \cref{eq:bcft_bulk_expansion}. The correction \eqref{eq:var_bcft_two_pt} can then be written as the sum of two terms,
\begin{align}
\label{eq:variation_bcft_two_pt}
\begin{split}
        \frac{\braket{\delta_{\xi}\left(\Phi(p,\bar{p})\overline{\Phi}(p',\bar{p}')\right)}_{\mathbb{H}^+}}{\braket{\Phi(p,\bar{p})\overline{\Phi}(p',\bar{p}')}_{\mathbb{H}^+}}  ={}d_n\mathrm{Re}\left(\xi'(p)+ \xi'(p')-\frac{2}{p-p'}(\xi(p)-\xi(p'))\right) \\
        +\partial_\eta \left(\log\left((1-\eta)^{d_n} K_\text{blk}(\eta)\right) \right)(\xi(p)\partial_{p}-\xi(\bar{p})\partial_{\bar{p}}+\xi(p')\partial_{p'}-\xi(\bar{p}')\partial_{\bar{p}'}) \eta.
    \end{split}
    \end{align} 
    The first term reflects the variation of $|p-p'|^{-2d_n}$ and the second term arises from the derivatives acting on the $\eta$-dependent pieces. Using the explicit form of $\xi$ given in \cref{eq:conf_trans} one finds
    \begin{equation}
    d_n\mathrm{Re}\left(\xi'(p)+ \xi'(p')-\frac{2}{p-p'}(\xi(p)-\xi(p'))\right)=2d_n\tilde{\lambda}_3(p_x+p_x')\left(1-\frac{1-\eta}{1-\eta}\right)=0,
\end{equation}
which must be the case, since $|p-p'|^{-2d_n}$ takes the form of a CFT two-point function which is invariant under all global conformal transformations.

In order to find full invariance of the entanglement entropy it is therefore necessary that 
\begin{align}
    \label{eq:condition}
    K_{\mathrm{blk}}(\eta) = \sigma (1-\eta)^{-d_n} + \mathcal O((n-1)^2),
\end{align}
for some constant $\sigma$, such that the second line of \cref{eq:variation_bcft_two_pt} vanishes faster than $(n-1)$ as $n \to 1$.

In fact, this condition is naturally satisfied in the large-$c$ limit. In this limit, conformal blocks exponentiate\footnote{Since we work on the Replica geometry, we have to multiply the central charge with $n$.} \cite{Zamolodchikov:1987avt, Belavin:1984vu, Hartman:2013mia}
\begin{equation}
    \mathscr{F}(c, \Delta_i, d_n/2\,|\,1-\eta)=\exp\left(-\frac{nc}{6}f\left(\frac{\Delta_i}{nc}, \frac{d_n}{2nc},1-\eta\right)\right), \label{large c}
\end{equation} 
with the so-called semi-classical block $f$. In the large-$c$ limit and a neighborhood around $\eta = 1$, the leading contribution for light operators, i.e., operators for which $\Delta_i \sim \mathcal O(1)$, is $f\left(0, \frac{d_n}{2nc},1-\eta\right)$. The semi-classical block satisfies a particular monodromy problem whose solution is given by\footnote{A constant of integration is fixed by matching to the known expansion of the semi-classical block in the limit $\eta \to 1$.} \cite{Hartman:2013mia, Sully:2020pza} \begin{equation}
    f\left(0, \frac{d_n}{2nc},1-\eta\right)=\frac{6d_n}{nc}\log(1-\eta) + \mathcal O((n-1)^2). \label{limitlarge}
\end{equation}
Plugging this into \cref{large c} reproduces \cref{eq:condition} up to the prefactor $\sigma$.

The constant $\sigma$ has be to such that the von Neumann entropy of single interval on the plane is reproduced. If \cref{eq:condition} holds, we find that \cref{eq:bcft_bulk_expansion} reads
\begin{align}
    \braket{\Phi_n(p,\bar{p})\overline{\Phi}_n(p',\bar{p}')}_{\mathbb{H}^+} = \frac{\sigma}{|p - p'|^{2d_n}} + \mathcal O((n-1)^2),
\end{align}
such that by comparing with \cref{eq:twist_two_pt_plane} we can identify
\begin{align}
    \label{eq:fixed_sigma}
    \log \sigma = \log \mathscr{C}_{\bar \Phi_n \Phi_n} + \text{(subleading)}.
\end{align}
The subleading corrections to the $\mathcal O(c)$ leading term are allowed, since we only study von Neumann entropy at leading order in the central charge.

Comparing to \cref{eq:bcft_bulk_expansion}, we see that $\sigma$ must arise from the sum over primaries.
Contributions from heavy primaries, i.e., operators for which $\Delta_i \gtrsim \mathcal O(c)$ are exponentially suppressed\footnote{In the next section we will mention subtleties for operators with $\Delta \sim \mathcal O(c)$.}, as long as we are close enough to $\eta = 1$. We can therefore restrict to contributions from light internal operators for which $\Delta_i/c \to 0$ in the large-$c$ limit such that their contribution is given by the vacuum block. As explained above, the yields the correct dependence on $\eta$ to reproduce \cref{eq:condition}. The constant $\sigma$ then becomes
\begin{align}
    \sigma = \mathscr{C}_{\Phi_n\overline{\Phi}_n} + \sum_{\Delta_i > 0}^{\Delta_i < \mathcal O(c)} \mathscr{C}_{\Phi_n\overline{\Phi}_n}^{\mathcal O_i}\mathscr{A}_i,
\end{align}
where we have written the vacuum contribution $\mathscr{C}_{\Phi_n\overline{\Phi}_n}^\mathbbm{1} = \mathscr{C}_{\Phi_n\overline{\Phi}_n}$ separately. The condition \eqref{eq:fixed_sigma} now requires that the sum over all coefficients except for the vacuum contribution must be smaller than $\mathcal O(\mathrm{e}^c)$. A similar constraint is found in holographic theories and typically leads to the requirement of a small number of low-dimension operators \cite{Hartman:2014oaa}. Here, however, we do not need to worry about the presence of too many light operators. Instead we can ask that the only non-vanishing one-point functions $\mathscr A_i$ are those of sufficiently few light operators. In fact, in the AdS$_2$/bath system with no additional sources turned on we expect that the identity is the only operator with a non-vanishing one-point function $\mathscr{A}_{1} = 1$. In this case $K_\text{blk}(\eta)$ reduces to
\begin{align}
\label{eq:K_at_large_c}
    K_\text{blk}(\eta) = \mathscr{C}_{\bar \Phi_n \Phi_n} \exp\left(-\frac{nc}{6}f\left(0, \frac{d_n}{2nc},1-\eta\right)\right)  \qquad \text{(at large $c$)},
\end{align}
independently of the details of the spectrum. Substituting \cref{limitlarge} into \cref{eq:K_at_large_c} exactly reproduces \cref{eq:condition}. 

This shows that the large-$c$ limit ensures that in a neighborhood around $\eta = 1$ the BCFT von Neumann entropy in the bulk channel in the presence of a deformed boundary is correctly reproduced from an AdS$_2$/bath computation, as long as \cref{eq:condition} holds together with the condition that $\mathscr A_i \neq 0$ for only a $\mathcal O(1)$ number of light operators.

\subsection{Boundary Channel}

\subsubsection{Entanglement Entropy in the BCFT}
We probe again the entanglement entropy of intervals $A$ ending at $p,p'\in\mathbb{H}^+$ under a infinitesimal boundary deformation, but now in the boundary channel. The relevant OPE channel in this case, i.e. the limit $\eta \to 0$, is obtained by first expanding each
operator using the BOE, so that the bulk two-point function becomes a sum of boundary
two-point functions.

In the boundary channel, the BCFT two point function is given by 
\begin{align}
    \braket{\Phi_n(p,\bar{p})\overline{\Phi}_n(p',\bar{p}')}_{\mathbb{H}^+} =\left(\frac{\eta}{4p_x p_x'}\right)^{d_n}\sum_I\mathscr{B}_{\Phi_n I}\mathscr{B}_{\overline{\Phi}_nI}\mathscr{F}(c,\hat{\Delta}_I,d_n/2\,|\,\eta). \label{4 point}
\end{align} 
again with the conformal cross ratio \eqref{eq:bcft_cross_ratio}. The sum now runs over boundary primaries and $\mathscr B_{iJ}$ are the BOE coefficients which arise when expanding a bulk operator $\mathcal O_{\Delta_i}$ in terms of boundary operators. 

We now act with the infinitesimal deformation on the BCFT. Similar as in the bulk channel, we introduce a short hand notation for the contributions of the blocks
\begin{equation}
    K_{\mathrm{bd}}(\eta)=\sum_I\mathscr{B}_{\Phi_n I}\mathscr{B}_{\overline{\Phi}_nI}\mathscr{F}(c,\hat{\Delta}_I,d_n/2\,|\,\eta),
\end{equation}
such that the infinitesimal transformation of \eqref{4 point} under boundary deformations can be written as 
\begin{align}
\label{eq:variation_two_pt_bdry_channel}
&\frac{\braket{\delta_\xi\left(\Phi_n(p,\bar{p})\overline{\Phi}_n(p',\bar{p}')\right)}_{\mathbb{H}^+}}{\braket{\Phi_n(p,\bar{p})\overline{\Phi}_n(p',\bar{p}')}_{\mathbb{H}^+}} \notag \\&=d_n\left(\mathrm{Re}\,\xi(p)-
\frac{1}{\mathrm{Im}\,p}\mathrm{Im}\,\xi(p) +\mathrm{Re}\,\xi(p')-
\frac{1}{\mathrm{Im}\,p'}\mathrm{Im}\,\xi(p')\right) \notag \\ &+ \partial_\eta \log \left( \eta^{d_n}K_{\mathrm{bd}}(\eta) \right)(\xi(p)\partial_{p}-\xi(\bar{p})\partial_{\bar{p}}+\xi(p')\partial_{p'}-\xi(\bar{p}')\partial_{\bar{p}'})\eta,
\end{align} 
where the first row arises from the variation of the term $(4p_xp_x')^{-d_n}$. However, unlike in the previous case, this expression does not vanish. The second row comes from the derivatives acting on the $\eta$-dependent pieces. As we are in the large-$c$ limit, the conformal blocks again exponentiate and for light operators in the boundary channel all blocks are equal to \cite{Sully:2020pza}
\begin{equation}
\label{eq:large_c_boundary_block_limit}
    \mathscr{F}(c, \hat{\Delta}_I, d_n/2\,|\,\eta)= \eta^{-d_n} \mathcal O((n-1)^2).
\end{equation}
Substituting this into \cref{eq:variation_two_pt_bdry_channel} the variation of the two-point function thus reduces to
\begin{align}
\label{eq:change_two_pt_function}
&\braket{\delta_\xi\left(\Phi_n(p,\bar{p})\overline{\Phi}_n(p',\bar{p}')\right)}_{\mathbb{H}^+}\notag \\&=d_n\left(\mathrm{Re}\,\xi(p)-
\frac{1}{\mathrm{Im}\,p}\mathrm{Im}\,\xi(p) +\mathrm{Re}\,\xi(p')-
\frac{1}{\mathrm{Im}\,p'}\mathrm{Im}\,\xi(p')\right)\braket{\Phi_n(p,\bar{p})\overline{\Phi}_n(p',\bar{p}')}_{\mathbb{H}^+}.
\end{align}
Interestingly, this expression is again valid even for a dense low-lying spectrum, since the multiplicity of light operators does not affect the cancellation in the second line of \cref{eq:variation_two_pt_bdry_channel}.
If we compare this expression to our result in the single-interval case, \cref{eq:change_one_pt_function}, we see that in the two-point case the factor multiplying the correlation function on the right hand side of \cref{eq:change_two_pt_function} is obtained by summing the corresponding term of the one-point case, \cref{eq:change_one_pt_function}, over each point. We might thus expect that when using this expression to evaluate the von Neumann entropy, we will obtain a sum over terms which take the form of dilaton profiles evaluated at certain locations.

And in fact,  the entanglement entropy in the undeformed case is
\begin{equation}
    S(A)=\frac{c}{6}\log 2p_x+\frac{c}{6}\log 2p_x'+\lim_{n\to 1}\frac{1}{1-n}\log\sum_I\mathscr{B}_{\Phi_n I}\mathscr{B}_{\overline{\Phi}_nI}
\end{equation}
and the first correction to the entanglement entropy is reads
\begin{align}
    \delta\tilde{S}(A)=-\frac{c}{6}\left(\mathrm{Re}\,\xi(p)-
\frac{1}{\mathrm{Im}\,p}\mathrm{Im}\,\xi(p) +\mathrm{Re}\,\xi(p')-
\frac{1}{\mathrm{Im}\,p'}\mathrm{Im}\,\xi(p')\right).
\end{align}
Plugging in the explicit form of the boundary deformation we arrive at
\begin{align}
    \tilde{S}(A)&=\frac{c}{6}\log 2p_x+\frac{c}{6}\log 2p_x'+\lim_{n\to 1}\frac{1}{1-n}\log\sum_I\mathscr{B}_{\Phi_n I}\mathscr{B}_{\overline{\Phi}_nI}\notag \\ &+\frac{c}{6}\left( \frac{\tilde{\lambda}_1+\tilde{\lambda}_2 p_\tau+\tilde{\lambda}_3(p_\tau^2+p_x^2)}{p_x}+\frac{\tilde{\lambda}_1+\tilde{\lambda}_2 p_\tau'+\tilde{\lambda}_3((p_\tau')^2+(p_x')^2)}{p_x'}\right). \label{eq: Ent final bd channel}
\end{align}

It is now useful to separate the vacuum contribution in the sum i.e. \begin{equation}
\label{eq:sum_b}
    \sum_I\mathscr{B}_{\Phi_n I}\mathscr{B}_{\overline{\Phi}_nI}=\mathscr{B}_{\Phi_n \mathbbm{1}}\mathscr{B}_{\overline{\Phi}_n \mathbbm{1}}+\sum_{I\neq \mathbbm{1}}\mathscr{B}_{\Phi_n I}\mathscr{B}_{\overline{\Phi}_nI}=\mathscr{A}_{\Phi_n}^2\sum_{I}\overline{\mathscr{B}}_{\Phi_n I}\overline{\mathscr{B}}_{\overline{\Phi}_nI}
\end{equation} where we used $\mathscr{A}_{\Phi_n}=\mathscr{B}_{\Phi_n \mathbbm{1}}$ and defined normalized BOE coefficients $\overline{\mathscr{B}}_{\mathcal O J} =\frac{\mathscr{B}_{\mathcal O J}}{\mathscr{B}_{\mathcal O \mathbbm{1}}}$ by dividing all coefficients by $\mathscr{B}_{\Phi_n \mathbbm{1}}$. As explained in \cite{Sully:2020pza} (see also \cite{Perlmutter:2013paa}) an overall factor $\mathscr{A}_{\Phi_n}$ is common to all OPE coefficients $\overline{\mathscr{B}}_{\Phi_n J}$ and so the dependence on $\epsilon$ and $g_b$ cancels for all terms in the sum.

Using \cref{A Def}, we find the expression
\begin{align}
    \tilde{S}(A)&=\frac{c}{6}\log \frac{2p_x}{\epsilon}+\frac{c}{6}\log \frac{2p_x'}{\epsilon}+2\log g_b+\lim_{n\to 1}\frac{1}{1-n}\log\sum_{I}\overline{\mathscr{B}}_{\Phi_n I}\overline{\mathscr{B}}_{\overline{\Phi}_nI}\notag \\ &+\frac{c}{6}\left( \frac{\tilde{\lambda}_1+\tilde{\lambda}_2 p_\tau+\tilde{\lambda}_3(p_\tau^2+p_x^2)}{p_x}+\frac{\tilde{\lambda}_1+\tilde{\lambda}_2 p_\tau'+\tilde{\lambda}_3((p_\tau')^2+(p_x')^2)}{p_x'}\right). \label{eq:BCFTMulti}
\end{align}
Note the coefficients $\mathscr{B}_{\Phi_n J}$ vanish for $J \neq 1$ as $n \to 1$. In fact, they vanish as $\mathscr{B}_{\Phi_n J} = \mathcal O(n-1)$ and so the the sum takes the form $\sum_{I}\overline{\mathscr{B}}_{\Phi_n I}\overline{\mathscr{B}}_{\overline{\Phi}_nI} = 1 + \mathcal O(n-1)$. As a result the remaining limit is well defined.

\subsubsection{Island Entropy of the AdS/Bath-System}
\begin{figure}[t]
    \centering
\def\svgwidth{7cm}
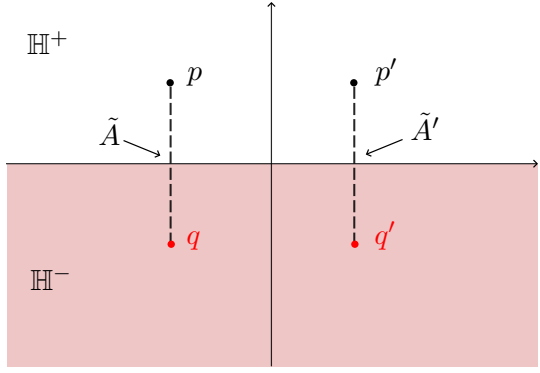
\caption{Two intervals $\tilde{A}$ and $\tilde{A}'$ in the AdS$_2$/bath geometry. The upper half plane is the non gravitating bath, the lower half plane is the AdS$_2$-region.}
\label{fig:AdS/Bath2}
\end{figure}
Similarly as for the single interval case, we would like to compare \eqref{eq: Ent final bd channel} to an island entropy in an AdS$_2$/bath system. The geometry is the same as in \cref{subsubsec: Island entropy of AdS bath system}. 
We introduce twist operators at two points $p,p'\in\mathbb{H}^+$ in the upper half plane and two points $q,q'\in\mathbb{H}^-$ in the lower half plane, see \cref{fig:AdS/Bath2}. Expanding in the channel indicated in \cref{fig:AdS/Bath2} we find the generalized entropy
\begin{align}
    S_\mathrm{gen}([p,q]\cup[p',q'])&=\frac{\mathrm{1}}{4G_\mathrm{N}}(2\phi_0+\phi(q)+\phi(q'))+\frac{c}{6}\log \frac{|p-q|^2}{-q_x}+\frac{c}{6}\log \frac{|p'-q'|^2}{- q_x'}\notag \\ &+\lim_{n\to 1}\frac{1}{1-n}\log\left(|\eta|^{2d_n}\sum_i\mathscr{C}^i_{\Phi_n\overline{\Phi_n}}\mathscr{C}^i_{\Phi_n\overline{\Phi_n} }\,|\mathscr{F}(c, \Delta_i, d_n/2\,\,|\,\eta)|^2\right), \label{ent block 2}
\end{align}
where as before $\mathscr{F}$ are chiral conformal blocks, $\mathscr{C}^i_{\Phi_n\overline{\Phi_n}}$ are the OPE coefficients from the twist field products and the numerators $q_x$ and $q_x'$ arise from the fact that $q, q'$ are located in the AdS$_2$-region.

We use again the island formula to determine the location of the points $q, q'$ in the gravitating region, following the same steps as before: we write down the generalized entropy for arbitrary $q,q'$ and extremize with respect to $q,q'$. In the large-$c$ limit, the problem simplifies dramatically. The conformal block exponentiates as before, c.f.\ \cref{eq:large_c_boundary_block_limit}, which removes the $\eta$-dependence in the last line of \cref{ent block 2}. Therefore, the extremization of $q$ and $q'$ can be done independently and each extremization reduces to the problem discussed in single interval case, \cref{subsubsec: Island entropy of AdS bath system}. We find 
\begin{equation}
    \tilde{q}=\bar{p}, \hspace{0.5cm} \tilde{q}'=\bar{p}'
\end{equation} as quantum extremal surfaces. Inserting this into \eqref{ent block 2} and turning on our non trivial dilaton profile yields
\begin{align}
    S_\mathrm{island}(p,p')&=\frac{\mathrm{1}}{4 G_\mathrm{N}}(2\phi_0+\phi(p)+\phi(p'))+\frac{c}{6}\log 2p_x+\frac{c}{6}\log 2 p_x'+2\cdot\frac{c}{6}\log 2\notag \\ &+\lim_{n\to 1}\frac{1}{1-n}\log\sum_i\mathscr{C}^i_{\Phi_n\overline{\Phi_n}}\mathscr{C}^i_{\Phi_n\overline{\Phi_n} }. \label{Island entropy multiple}
\end{align}
Similar to the BCFT computation, we can separate the identity contribution to the sum over primaries, $\overline{\mathscr{C}}^i_{\Phi_{n}\overline{\Phi_{n}}}=\frac{\mathscr{C}^i_{\Phi_{n}\overline{\Phi_{n}}} }{\mathscr{C}^{\mathbbm 1}_{\Phi_{n}\overline{\Phi_{n}}} }$, and use the identity $\lim_{n\to 1}\frac{1}{1-n}\log\mathscr{C}_{\Phi_n\overline{\Phi}_n}^{\mathbbm 1}=-\frac{c}{3}\log\epsilon$ to obtain 
\begin{align}
    S_\mathrm{island}(p,p')&=\frac{\mathrm{1}}{4 G_\mathrm{N}}(2\phi_0+\phi(p)+\phi(p'))+\frac{c}{6}\log \frac{2p_x}{\epsilon}+\frac{c}{6}\log \frac{2 p_x'}{\epsilon}+\frac{c}{3}\log \frac{2}{\epsilon}\notag \\ &+\lim_{n\to 1}\frac{1}{1-n}\log\sum_{i}\overline{\mathscr{C}}^i_{\Phi_n\overline{\Phi_n}}\overline{\mathscr{C}}^i_{\Phi_n\overline{\Phi_n} }.
\end{align}

\subsubsection{Matching with BCFT Calculation in the Boundary Channel}
We entered our computation with the expectation that we can match the entropies from the BCFT and AdS$_2$/bath computation if we impose large central charge and assume the cross ratio to be near $\eta\to 0$. As before, using \cref{eq:gnewton} and the explicit from of the dilaton solution, \cref{dilaton}, the dilaton contribution matches with the correction coming from the boundary deformation in \cref{eq: Ent final bd channel}, i.e., 
\begin{align}
 \lambda_i=\tilde{\lambda}_i.
\end{align}
However, the leading order entropy only agrees with \cref{eq: Ent final bd channel} if
\begin{equation}
    \label{eq:OPE_condition}
     \lim_{n\to 1}\frac{1}{1-n}\log\sum_I\overline{\mathscr{B}}_{\Phi_n I}\overline{\mathscr{B}}_{\overline{\Phi}_nI}=\lim_{n\to 1}\frac{1}{1-n}\log\sum_i\overline{\mathscr{C}}^i_{\Phi_n\overline{\Phi_n}}\overline{\mathscr{C}}^i_{\Phi_n\overline{\Phi_n} }
 \end{equation} 
 at leading order in $c$.
 These contributions to the entanglement entropy are finite and were computed in a number of different scenarios e.g.\ in \cite{Calabrese:2010he,Perlmutter:2015iya,Sully:2020pza}. In the most general case, light operators of course also will add a non-trivial dependence on $\eta$ to the entanglement entropy, which however is subleading in $c$.

 For holographic theories the sparse spectrum condition ensures that both sides of \cref{eq:OPE_condition} are of $\mathcal O(1)$ and in this case both the BCFT and AdS$_2$/bath computations agree. If we do not want to assume that the BCFT is holographic, \cref{eq:OPE_condition} places a new condition on the theories for which the effect of boundary deformations on entanglement entropy can be computed from the island formula in JT gravity.

\subsection{Conditions on the Spectrum}
So far we have seen that near the locations $\eta = 1$ and $\eta = 0$, where vacuum block-type contributions in the bulk and boundary channel, respectively, dominate, the BCFT computation agrees with a computation performed in an AdS$_2$/bath system in the large-$c$ limit, if the BCFT is holographic. If it is not holographic, there will in general be additional constraints on the OPE coefficients as just explained. The agreement holds true in a finite neighborhood $[0, \eta^*_0]$ and $[\eta^*_1, 1]$ around those points. As we cross $\eta^*_i$, either a non-vacuum block starts to dominate, or the contribution of many (heavy) primaries have to be resummed and it is favorable to go to expand the correlation function in a different channel where again the vacuum blocks originating from light operators dominate. Both effects are non-perturbative in $1/c$ and therefore cannot be seen in our analysis. Our computation requires that the vacuum block dominates in the channel which gives the most important contribution. Therefore, we require that no contribution coming from non-vacuum blocks should dominate the sum over blocks and $\eta_0^* = \eta_1^*$. A more detailed study under which condition this happens is beyond the scope of this work and we leave it for the future.

This situation is well known from the study of entanglement entropies in two-dimensional CFTs at large central charge. For two-dimensional holographic CFTs requirements on the spectrum and OPE data to reproduce a gravitational computation were analyzed in \cite{Hartman:2013mia,Hartman:2014oaa}. For the case of a holographic BCFT, this was nicely done in \cite{Sully:2020pza}. However, here, we do not want to assume that our BCFT is holographic. In particular it does not have to be dual to a theory with an $\mathcal O(1)$ number of light fields. We can therefore relax some of the assumptions that are typically made for holographic (B)CFTs.

The statement that only the vacuum block appears at large $c$ is known as \emph{vacuum block dominance}. The authors of \cite{Sully:2020pza} defined it as follows:

\begin{definition}[Vacuum Block Dominance]
\label{def:vac_dom}
\hfill
\begin{enumerate}
    \item The contribution of heavy operators in neighborhoods around $\eta = 0,1$ is exponentially suppressed in $c$.
    \item Contributions from non-vacuum operators to the coefficient of the vacuum block must be subleading in $c$ with respect to the vacuum, e.g., the sums over $\overline{\mathscr{B}}_{\overline{\Phi}_nI}$ or $\overline{\mathscr{C}}^i_{\Phi_{n}}$ are subexponential in $c$.
\end{enumerate}
\end{definition}

The first condition essentially ensures that the dominant block has the functional form of the vacuum block.
As discussed in section 4.4 of \cite{Sully:2020pza}, for operators whose dimensions scale as $\Delta \sim \mathcal O(c^\alpha)$ with $\alpha > 1$, the first point holds \cite{Pappadopulo:2012jk}. Moreover, the first requirement implies that the product of operator density and OPE coefficients for operators with dimension  $\Delta \sim \mathcal O(c)$ does not grow too fast,
\begin{align}
    \label{eq:intermediate_bound}
    \log\left(\rho(\delta) |B(\delta)|^2 \right) \leq c \delta \log \eta^{-1},
\end{align}
where $\rho(\delta)$ is the operator density and $B(\delta)$ is an averaged OPE coefficient for operators with $\Delta \in c [\delta, \delta + \mathrm{d}\delta]$ for some $\delta \sim \mathcal O(1)$ and $\mathrm{d}\delta \ll 1$. Our computation relies in a similar way on vacuum block dominance, and so the first point in \cref{def:vac_dom} must also hold in our case. 

The second point in the definition would require the contribution on both sides of \cref{eq:OPE_condition} to be subleading in a large-$c$ expansion. In \cite{Sully:2020pza} this requirement was necessary, since otherwise it would yield an $\mathcal O(c)$ contribution to the entanglement entropy which is not reproduced by the gravitational computation in a higher-dimensional bulk. In the case of Einstein gravity, there is no $\log c$ contribution to the entropy and so the bound is even stronger and both sides of \cref{eq:OPE_condition} have to be $\mathcal O(1)$.

However, in our case we do not compare to a holographic computation in a higher-dimensional bulk, but rather to an AdS$_2$/bath computation in the same number of dimensions. In particular, the matter theory is the same in both cases. We thus do not need to assume that both sides of \cref{eq:OPE_condition} are of $\mathcal O(1)$, but it is sufficient to require that \cref{eq:OPE_condition} holds to leading order in $c$. Thus, in our case, a \emph{weaker form of vacuum block dominance} seems sufficient, namely

\begin{definition}[Weak Vacuum Block Dominance]
\label{def:weak_vac_dom}
\hfill
\begin{enumerate}
    \item The contribution of heavy operators in neighborhoods around $\eta = 0,1$ is exponentially suppressed in $c$.
    \item $\mathcal O(c)$ corrections to leading order entanglement entropy in the BCFT description must agree with corrections in the AdS$_2$/bath description.
\end{enumerate}
\end{definition}
It is in this sense that our derivation also works for non-holographic theories, i.e., we propose that at least the second point in the definition of vacuum block dominance can be relaxed compared to holographic theories. In its current form, the second point in the definition of weak vacuum block dominance is not very useful. To really understand the extent to which those conditions are universally true at large $c$ is an interesting questions which we leave for future work. Instead, in the upcoming section, we will substantiate its meaning by deriving a large set of conditions which need to hold in the many-interval case.

From the AdS$_2$/bath computation it is obvious that the presence of such $\mathcal O(c)$ corrections will change the value $\eta^*$ for which the entanglement entropy undergoes a phase transition between the phase without an island and with an island. The reason is that an $\mathcal O(\mathrm{e}^c)$ number of light operators produce a positive contribution only to the entropy in the island phase. Since the smaller of island and non-island value gives the correct entropy, the phase transition happens at a different value of $\eta$. Equivalently, it will affect the value of $\eta$ at which the vacuum block dominates the bulk or boundary channel. This of course can also be understood from a BCFT computation where the presence of a large number of light fields modifies the leading term in the boundary channel as
\begin{align}
    \exp\left(- \frac c 6 f_0(\eta) \right) \to \exp\left(- \frac c 6 f_0(\eta) + \alpha c\right), \hspace{0.4cm} \text{with} \hspace{0.4cm} \alpha = \frac 1 c \log \sum_I\overline{\mathscr{B}}_{\Phi_n I}\overline{\mathscr{B}}_{\overline{\Phi}_nI}.
\end{align}

\section{Multiple Intervals}
\label{sec:multiple_intervals_2}
We now briefly comment on the generalization to multiple intervals, which follows analogous results in holographic theories \cite{Hartman:2013mia,Sully:2020pza}. For the BCFT discussion involving $s$ intervals, we will fix $2s$ points in $\mathbb{H}^+$ and evaluate the BCFT $2s$-point function to compute the entanglement entropy via analytic continuation in $n$. As we did in our previous analysis we will then link the correction to the von Neumann entropy due to a boundary deformation to the dilaton contribution in the island entropy in the AdS$_2$/bath system.

\subsection{BCFT Calculation for Multiple Intervals}
The correlation function of interest is given by a product of $s$ twist and $s$ anti-twist operators. 
There are various choices of performing the bulk OPE or BOE of twist operators, analogous to the two-interval case. Each choice defines a fusion channel $\mathcal{E}$, which we will keep fixed. Each fusion channel comes with a set of cross ratios which we summarize in a vector $\vec{\eta}$, a set of internal weights $\vec{\Delta}$ and moreover depends on a set of OPE/BOE coefficients.

\begin{figure}[t]
    \centering
\def\svgwidth{13.5cm}
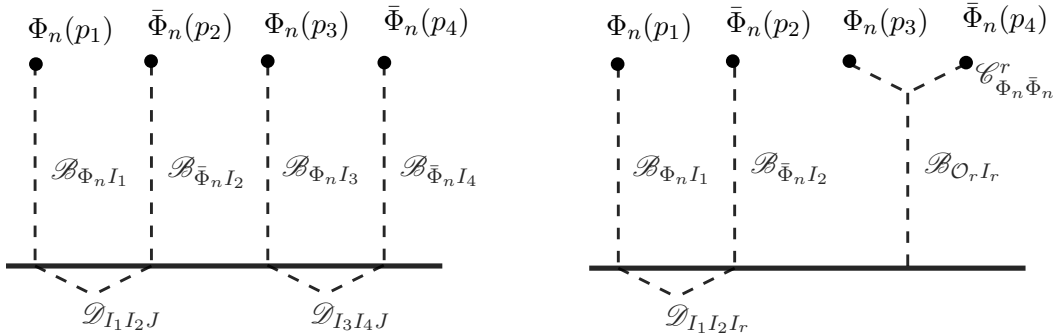
\caption{Demonstration of two different channels for a BCFT four-point function and its OPE coefficients.}
\label{fig:4Channel}
\end{figure}

To give a concrete example for the dependence on OPE coefficients, let us compare a BCFT four point function of twist operators \begin{equation}
\braket{\Phi_n(p_1, \bar{p}_1)\bar{\Phi}_n(p_2, \bar{p}_2)\Phi_n(p_3, \bar{p}_3)\bar{\Phi}_n(p_4, \bar{p}_4)}_{\mathbb{H}^+}
\end{equation} in two different channels: 
\begin{enumerate}
    \item We take the BOE of all operators first which we write schematically, dropping all coordinate dependence, as\begin{equation}
    \Phi_n\thicksim \sum_{I_i} \mathscr{B}_{\Phi_n I_i}\hat{\mathcal{O}}_{I_i}
\end{equation} with $i=1, \dots, 4$ and boundary operators $\hat{\mathcal{O}}_{I_i}$. The exact form is given in \eqref{eq:bulk_boundary_ope}. 
The fusion of $\hat{\mathcal{O}}_{I_1}$ with $\hat{\mathcal{O}}_{I_2}$ and $\hat{\mathcal{O}}_{I_3}$ with $\hat{\mathcal{O}}_{I_4}$ gives two more BOPE coefficients $\mathscr{D}_{I_1I_2 J}$ and $\mathscr{D}_{I_1I_2 K}$, respectively. Since a two point function of two boundary operators takes the simple form \begin{equation}
    \braket{\hat{\mathcal{O}}_J(\tau_J)\hat{\mathcal{O}}_K(\tau_K)}=\frac{\delta_{JK}}{|\tau_J-\tau_K|^{2\hat{\Delta}_J}},
\end{equation} we do not generate more OPE coefficients. Thus, the channel depends on the OPE/BOE coefficients
\begin{equation}
    \label{eq:channel_1_ope}
   \mathscr{B}_{\Phi_n I_1}, \hspace{0.3cm} \mathscr{B}_{\bar{\Phi}_n I_2}, \hspace{0.3cm} \mathscr{B}_{\Phi_n I_3}, \hspace{0.3cm} \mathscr{B}_{\bar{\Phi}_n I_4}, \hspace{0.3cm} \mathscr{D}_{I_1I_2 J}, \hspace{0.3cm} \mathscr{D}_{I_3I_4 J}.
\end{equation}

\item We choose the BOE to expand the twist operators at $p_1$ and $p_2$, but the bulk OPE to expand the operators at $p_3$ and $p_4$. This yields two BOE coefficients $\mathscr{B}_{\Phi_n I_1}$ and $\mathscr{B}_{\bar{\Phi}_n I_2}$ and one bulk OPE coefficient, since
\begin{equation}
    \Phi_n\bar{\Phi}_n\thicksim \sum_r \mathscr{C}_{\Phi_{n}\bar{\Phi}_{n}}^r\mathcal{O}_r
\end{equation} with internal bulk operators $\mathcal{O}_r$. The $\mathcal{O}_r$ can then be expanded in terms of boundary operators, giving a BOE coefficient $\mathscr{B}_{\mathcal{O}_r I_r}$. Lastly, we obtain one more boundary OPE coefficient $\mathscr{D}_{I_1 I_2 I_r}$ from the boundary operator three-point function. Thus, in this channel the relevant OPE/BOE coefficients are
\begin{equation}
    \label{eq:channel_2_ope}
   \mathscr{B}_{\Phi_n I_1}, \hspace{0.3cm} \mathscr{B}_{\bar{\Phi}_n I_2}, \hspace{0.3cm} \mathscr{C}_{\Phi_{n}\bar{\Phi}_{n}}^r, \hspace{0.3cm} \mathscr{B}_{\mathcal{O}_r I_r}, \hspace{0.3cm} \mathscr{D}_{I_1I_2 I_r}.
\end{equation}
\end{enumerate}
Figure \ref{fig:4Channel} summarizes our discussion.

We now return to our discussion of the $2s$-point function. In the large-$c$ limit the correlation function can be written as 
\begin{equation}
    \Big\langle\prod_{j=1}^s\Phi_n(p_{2j-1},\bar{p}_{2j-1})\overline{\Phi}_n(p_{2j},\bar{p}_{2j})\Big\rangle_{\mathbb{H}^+}=\sum_\mathscr{I}\mathscr{M}^{\mathcal{E},\mathscr{I}}\exp\left(-\frac{nc}{6}f^\mathcal{E}\left(\frac{\vec{\Delta}}{nc},\frac{d_n}{2nc}, p_j\right)\right).
\end{equation} 
The sum runs over all internal operators, labeled by the index set $\mathscr{I}$. In the above discussed four-point function the index set is $\mathscr{I}=\{I_1, I_2, I_3, I_4, J\}$ in the first case and $\mathscr{I}=\{I_1, I_2, r, I_r\}$ in the second case. The multi-OPE-coefficient $\mathscr{M}^{\mathcal{E},\mathscr{I}}$ is the product of all entering OPE coefficients in the channel expansion, such as \cref{eq:channel_1_ope,eq:channel_2_ope}, and $f^\mathcal{E}$ is the semiclassical block. 

As before, in a suitable kinematic regime only light operators dominate the sum and in the large-$c$ limit the semi-classical block can be replaced by the vacuum block $f^\mathcal{E}(\vec 0, \frac{d_n}{2nc},p_j)$ associated to the channel $\mathcal E$. The exact nature of the vacuum block in the case of many twists is slightly subtle and has been discussed by Hartmann \cite{Hartman:2013mia}:
In all branches of a channel where an external operator (i.e.\ twist) does not meet two internal operators, we pick out only the vacuum contribution of the internal operators. If it meets two internal operators one of them is chosen to be the identity and the other one has the same dimension as the external one. To be precise, by vacuum contribution we mean that the structure of the block is that of a block where the vacuum operator runs on an internal line. However, the (multi-)OPE coefficients are summed over all light operators as before and here light operators are not replaced by the vacuum coefficients.

For a given channel $\mathcal{E}$ in the large-$c$ limit and an appropriate limit of cross ratios $\vec{\eta}\to\vec{\eta}_0$, the leading contribution of $f^\mathcal{E}$ for light operators can again be determined by solving the monodromy problem. This is doable in the limit $n\to 1$, where the monodromy problem breaks down to sum of independent differential equations. One finds \cite{Sully:2020pza} \begin{equation}
    f^\mathcal{E}\left(\vec{0}, \frac{d_n}{2nc}, p_i\right)=(n-1)\left(\sum_{k\ell}\log |p_{k\ell}|^2+\sum_{m} \log p_{mm^*}\right)+\mathcal{O}((n-1)^2)
\end{equation} with $p_k-p_\ell=p_{k\ell}$ and $|p_m-\bar{p}_m|=p_{mm^*}$. We have chosen our notation to indicate which expansion is used for the twist at points $p_k$. Indices $k,\ell$ label the pairing of twist and anti-twist operators in the upper half plane $\mathbb{H}^+$, giving bulk OPE coefficients. The index $m$ labels the fusing of twist fields with the boundary, giving boundary OPE coefficients. 
The full correlation function then takes the form \begin{equation}
    \Big\langle\prod_{j=1}^s\Phi_n(p_{2j-1},\bar{p}_{2j-1})\overline{\Phi}_n(p_{2j},\bar{p}_{2j})\Big\rangle_{\mathbb{H}^+}=\sigma\exp\left(-\frac{nc}{6}f^\mathcal{E}\left(\vec{0},\frac{d_n}{2nc}, p_j\right)\right)
\end{equation}
with
\begin{equation}
    \sigma=\mathscr{M}^{\mathcal{E},\mathbbm{1}}+\sum_{\mathscr{I}\neq \mathbbm{1}}\mathscr{M}^{\mathcal{E},\mathscr{I}}=\underbrace{\mathscr{A}_{\Phi_n}^M(\mathscr{C}_{\Phi_n\overline{\Phi}_n}^{\mathbbm{1}})^N}_{\mathscr{M}^{\mathcal{E},\mathbbm{1}}}\sum_{\mathscr{I}}\overline{\mathscr{M}}^{\mathcal{E},\mathscr{I}} ,\label{sigmamulti}
\end{equation} where $M$ and $N$ are the numbers of BOE and bulk OPE coefficients involving the external twists. Similar as before, we have introduced $\overline{\mathscr{M}}^{\mathcal{E},\mathscr{I}}=\frac{\mathscr{M}^{\mathcal{E},\mathscr{I}}}{\mathscr{M}^{\mathcal{E},\mathbbm{1}}}$. Using again $\mathscr{A}_{\Phi_n}=g_b^{(1-n)}\epsilon^{d_n}$ and $\mathscr{C}_{\Phi_n\overline{\Phi}_n}^{\mathbbm{1}}=\epsilon^{2d_n}$ gives us the entanglement entropy \begin{equation}
    \label{eq:multi_ee}
    S(A)=\sum_{k\ell}\frac{c}{3}\log \frac{|p_{k\ell}|}{\epsilon}+\sum_m\left(\frac{c}{6}\log \frac{p_{mm^*}}{\epsilon}+\log g_b\right)+\lim_{n\to 1}\frac{1}{1-n}\log\sum_{\mathscr{I}}\overline{\mathscr{M}}^{\mathcal{E},\mathscr{I}}.
\end{equation} 

Finally, we consider again an infinitesimal deformation of the boundary by $\xi$ given in equation \eqref{eq:conf_trans}.
The first sum in \cref{eq:multi_ee} arises from the part $|p_k-p_\ell|^{-2d_n}$ of the correlation function, which is invariant under global conformal transformations. The second sum arises from the $|p_m-\bar{p}_m|^{-d_n}$-part of the correlation function. Its transformation behavior is already known from \cref{sec:multiple_intervals}, so the correction becomes
\begin{equation}
    \delta\tilde{S}(A)=\frac{c}{6}\sum_m\frac{\tilde{\lambda}_1+\tilde{\lambda}_2 p_{m,\tau}+\tilde{\lambda}_3(p_{m,\tau}^2+p_{m,x}^2)}{p_{m,x}},
\end{equation} i.e., a sum of single point boundary deformation. This is analogously to the BCFT one interval case in the boundary channel, such that the whole expression for the entanglement entropy in the channel $\mathcal E$ reads 
\begin{align}
    \label{eq:bcft_multi}
    \tilde{S}(A)&=\sum_{k\ell}\frac{c}{3}\log \frac{|p_{k\ell}|}{\epsilon}+\sum_m\left(\frac{c}{6}\log \frac{p_{mm^*}}{\epsilon}+\log g_b\right)+\lim_{n\to 1}\frac{1}{1-n}\log\sum_{\mathscr{I}}\overline{\mathscr{M}}^{\mathcal{E},\mathscr{I}} \notag \\ &+\frac{c}{6}\sum_m\frac{\tilde{\lambda}_1+\tilde{\lambda}_2 p_{m,\tau}+\tilde{\lambda}_3(p_{m,\tau}^2+p_{m,x}^2)}{p_{m,x}}.
\end{align} 

\subsection{Island Entropy of the AdS/Bath-System}

\begin{figure}[t]
    \centering
\def\svgwidth{13.5cm}
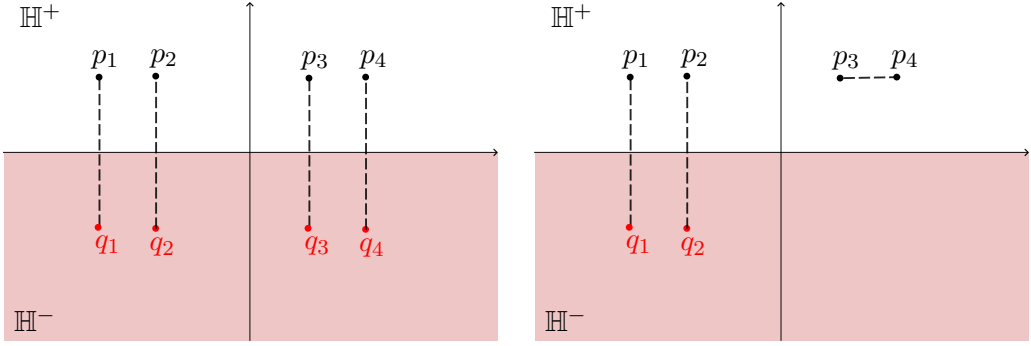
\caption{Channels in the AdS$_2$/bath system associated to the channels indicated in \cref{fig:4Channel}.}
\label{fig:AdSBath3}
\end{figure}

To match with the island entropy in the AdS$_2$/bath system, we fix $2s$ points $p_1, \dots, p_{2s}$ in the upper half plane. Now, each channel $\mathcal{E}$ in the BCFT can be associated to a channel in the AdS$_2$/bath system in the following way: For each twist in the BCFT that is expanded using the BOE we introduce a twist at a point $q_i$ in the gravitating region, i.e., in the lower half plane. The associated CFT channel is then defined by making sure that the operators at $p_i$ and $q_i$ are fused in the OPE.
\Cref{fig:AdSBath3} indicates the CFT channels associated to the respective channels in a BCFT shown in \cref{fig:4Channel}. 
For any such channel the computation runs in parallel to the two-interval case. In the large-$c$ limit, the conformal blocks exponentiate and the semiclassical block takes a closed expression near $\vec{\eta}\to\vec{\eta}_0$. We denote by $A$ the union of all intervals in our AdS$_2$/bath system.

We obtain for the generalized entropy \begin{align}
    S_\mathrm{gen}(A)&=\sum_m\left(\frac{1}{4 G_\mathrm{N}}(\phi_0+\phi(q_m))+\frac{c}{6}\log\frac{|p_m-q_m|^2}{-q_{mx}}\right)+\sum_{k\ell}\frac{c}{3}\log |p_{k\ell}| \notag \\ &+\lim_{n\to 1}\frac{1}{1-n}\log\sum_{\mathscr{J}}\mathscr{M}_\mathbb{C}^{\mathcal{E},\mathscr{J}}.
\end{align} 
Only points $q_m$ in the AdS$_2$ regions have dilaton contributions in the generalized entropy.
The index $m$ runs over twists in the bath region which are fused with twists in the AdS$_2$-region. These will give island contributions to the entropy. The index $k\ell$ runs over pairs of twists which are expanded using the OPE and will not lead to islands. Similarly to the BCFT case the multi-OPE-coefficient $\mathscr{M}_\mathbb{C}^{\mathcal{E},\mathscr{J}}$ is a product of all OPE coefficients which arise in the selected channel. The sum over internal operators is again captured by a sum over a multi-index $\mathscr{J}$. We do not have a boundary and therefore only bulk OPE coefficients enter. 

We define the vacuum multi-OPE coefficient as \begin{equation}
\mathscr{M}_\mathbb{C}^{\mathcal{E},\mathbbm{1}}=\prod_m\mathscr{C}^\mathbbm{1}_{\Phi_{n}\overline{\Phi}_{n}}\prod_{k\ell}\mathscr{C}^\mathbbm{1}_{\Phi_{n}\overline{\Phi}_{n}},
\end{equation}
where the indices follow the convention just explained.

We can now compute the island entropy. The extremization with respect to the $q_m$ are all independent and we obtain \begin{equation}
    \tilde{q}_m=\overline{p}_m,
\end{equation} which gives the island entropy (with dilaton profile \eqref{dilaton}  turned on) \begin{align}
    S_\mathrm{island}(p_1, \dots, p_m, p_{k\ell})&=\sum_m\left(\frac{1}{4 G_\mathrm{N}}(\phi_0+\phi(p_m))+\frac{c}{6}\log p_{mm^*}\right)+\sum_{k\ell}\frac{c}{3}\log |p_{k\ell}| \notag\\&+M\frac{c}{6}\log 2 +\lim_{n\to 1}\frac{1}{1-n}\log\sum_{\mathscr{J}}\mathscr{M}_\mathbb{C}^{\mathcal{E},\mathscr{J}}.
\end{align} Separating again the identity operator contribution from the sum $\sum_{\mathscr{J}}$, we arrive at \begin{align}
    S_\mathrm{island}(p_1, \dots, p_{2s})&=\sum_m\left(\frac{1}{4 G_\mathrm{N}}(\phi_0+\phi(p_m))+\frac{c}{6}\log\frac{p_{mm^*}}{\epsilon}\right)+\sum_{k\ell}\frac{c}{3}\log\frac{|p_{k\ell}|}{\epsilon} \notag\\& +M\frac{c}{6}\log \frac{2}{\epsilon} +\lim_{n\to 1}\frac{1}{1-n}\log\sum_{\mathscr{J}}\overline{\mathscr{M}}_\mathbb{C}^{\mathcal{E},\mathscr{J}}. \label{eq:IslandMulti}
\end{align} $M$ is again the number of fusions $p_m$ with $q_m$, i.e. AdS$_2$ to bath fusions.

\subsection{Matching of BCFT  and Island Calculation}
As before we see that $G_\mathrm{N}=\frac{3}{2c}$ and \eqref{dilaton} imply that the corrections in \cref{eq:bcft_multi} and \cref{eq:IslandMulti} agree under the identification $\lambda_i=\tilde{\lambda_i}$.
The agreement here is between individual channels in the BCFT and CFT computations and thus the final entanglement entropy will also agree.\footnote{This is at least as long as there is not an additional CFT channel that dominates and is was not considered. Understanding if this is the case would require more detailed knowledge of the conformal blocks at finite $c$.}

The leading piece of the entanglement entropy matches if \begin{equation}
    \lim_{n\to 1}\frac{1}{1-n}\log\sum_{\mathscr{I}}\overline{\mathscr{M}}^{\mathcal{E},\mathscr{I}}=\lim_{n\to 1}\frac{1}{1-n}\log\sum_{\mathscr{J}}\overline{\mathscr{M}}_\mathbb{C}^{\mathcal{E},\mathscr{J}}. \label{eq:Multicoeff}
\end{equation}
For holographic (B)CFTs both side of \eqref{eq:Multicoeff} are of order $\mathcal{O}(1)$ thanks to the sparse spectrum condition and agreement is guaranteed to the order we are interested in. Again, for non-holographic BCFTs, \cref{eq:Multicoeff} should be seen as a constraint on the theory for which BCFT boundary deformations can be modeled by a gravitating region with a non-trivial dilaton.

\section{Conclusions and Further Directions}
\label{sec:discussion}
In this paper we demonstrated that corrections to BCFT$_2$ entanglement entropy at large central charge in the presence of a global $\text{PSL}(2,\mathbb R)$ boundary deformation can, at leading order in the deformation, alternatively be computed as dilaton contributions to the island entropy of a BCFT with an undeformed boundary. In this latter case the (B)CFT is coupled to a gravitating region governed by JT gravity where the dilaton boundary condition is set by the original BCFT boundary deformation and the constant piece of the dilaton is determined in terms of the boundary $g$-function, \cref{eq:deformation_expectation,Bound entrop}.

We first showed this for the case of a single interval at zero as well as finite temperature, the latter situations being related to the former by a conformal transformation. These cases are completely fixed by symmetry. To provide further evidence for our claim, we therefore considered the case of multiple intervals at zero temperature which depends on the full BCFT spectrum. In order for the full entropy, including the leading piece to agree we needed to impose certain conditions on the spectrum and OPE coefficients of the BCFT, which are automatically fulfilled for holographic theories.

In particular, the number of operators with dimension $\mathcal O(c)$ must not grow too quickly, \cref{eq:intermediate_bound}, there must not be an $\mathcal O(\mathrm{e}^c)$ contribution coming from one-point functions in the bulk channel, see the discussion around \cref{eq:K_at_large_c}, and a certain relation between bulk and bulk-to-boundary OPE coefficients must hold, \cref{eq:OPE_condition,eq:Multicoeff}, at leading order in $c$. These conditions still imply that the kinematics of twist correlators are determined by the vacuum block, but are in general weaker than conditions for holographic two-dimensional CFTs. Inspired by the notion of vacuum block dominance which is satisfied by holographic theories, we say that theories obeying our conditions satisfy \emph{weak vacuum block dominance}, c.f.\ \cref{def:weak_vac_dom}.

There are several interesting directions in which our results can be taken.

Clearly, it is crucial to understand if the ``duality'' found in our paper extends beyond entanglement entropy. Obvious next targets would be the partition function at large $c$ and corrections subleading in $c$ as well as a study of local operator correlation functions both in the BCFT setup and in the AdS$_2$/bath setup with the same insertion points as in the BCFT case. Such studies would furthermore reveal if the spectrum conditions put forward in this paper are sufficient. Extending the current analysis to finite replica index $n$ would also be of interest. In this case, a main obstacle is to find the correct form of \emph{generalized R\'enyi entropies}, in particular the finite-$n$ generalization of the dilaton contribution. A candidate formula for generalized R\'enyi entropies has been put forward in \cite{Hollowood:2024uuf} and applying it to boundary deformations of large-$c$ BCFTs could serve as an important test of our and their proposals. A direct derivation in the case of a doubly-holographic model, where leading-order ETW brane-deformations are interpreted as a non-trivial dilaton profile on the undeformed brane worldvolume, would also be interesting directions for the future. Naturally, similar investigations can be done for other quantities such as entanglement asymmetry \cite{Ares:2022koq,Fossati:2024ekt, Benini:2024xjv}.

Our paper can be understood as providing evidence for a conjecture put forward in \cite{Neuenfeld:2024gta}. There, as explained in the introduction, the authors studied a doubly-holographic setup and showed that, at leading order, the deformations of an ETW brane are described by the dilaton of JT gravity on the brane which is consistent with our findings. Moreover, \cite{Neuenfeld:2024gta} also conjectured that for finite deformations the brane theory should become Liouville Theory with a dynamical background metric and matter fields. It would certainly be interesting to understand if this latter statement can be proven from the BCFT perspective alone.

It is well known that JT gravity has a boundary description as the Schwarzian \cite{Almheiri:2014cka} and therefore there should be a more direct relation between general boundary deformations in BCFT and the Schwarzian. This also pertains more general questions of dynamics. Results on the response of ETW branes in a three-dimensional bulk on to shocks is to appear soon \cite{QDF} and we expect those results to have a natural interpretation in terms of a BCFT as well as an AdS$_2$/bath system via double holography. It would then be natural to understand if one can again generalize those results to non-holographic BCFTs. Such a line of inquiry should connect to moving mirror models \cite{Akal:2021foz} in an interesting way.

Moreover, it would particularly be interesting to understand better the conditions on the BCFT under which our results hold and how to relax them. For example, so far we do not know if \cref{eq:OPE_condition} holds generically at large central charge, if it requires fine-tuning, or if it is perhaps impossible. One possible approach might be to study the BCFT crossing kernel for a heavy-heavy correlator at large central charge in order to find a universal relation between boundary OPE coefficients and bulk OPE coefficients along the lines of \cite{Numasawa:2022cni}. 

The arguably most speculative, but also most interesting, aspect of our work is that it suggest that boundary deformations of two-dimensional BCFTs can be understood in terms of a theory of gravity. Possibly related results in which entanglement dynamics of a BCFT is controlled by JT gravity appeared in \cite{Callebaut:2018nlq,Callebaut:2018xfu}, although the precise relation to our work still needs to be clarified. In our case the gravitational theory is semi-classical which is a result of the large-$c$ limit. If we were able to relax this strict limit, one might hope to obtain insights into properties of two-dimensional quantum gravity away from the semi-classical limit via the study of BCFT$_2$ boundary deformations. A first and more modest goal would be to relax the condition on the growths of operators of dimension $\mathcal O(c)$. This would result in other blocks besides the vacuum block being dominant and it would be extremely interesting to understand how such a result can be reproduced from a gravitational theory.

We hope to report progress on some of these questions in the future.

\acknowledgments

We thank Jan de Boer, Jildou Hollander, Jani Kastikainen, Thomas Kögel, Quim Llorens, Watse Sybesma, Andy Svesko, Frederik Taube for interesting discussions and collaboration on related topics.

\appendix

\bibliographystyle{JHEP}
\bibliography{refs}

@article{Ryu:2006bv,
    author = "Ryu, Shinsei and Takayanagi, Tadashi",
    title = "{Holographic derivation of entanglement entropy from AdS/CFT}",
    eprint = "hep-th/0603001",
    archivePrefix = "arXiv",
    reportNumber = "NSF-KITP-06-11, NSF-KITP-06-11",
    doi = "10.1103/PhysRevLett.96.181602",
    journal = "Phys. Rev. Lett.",
    volume = "96",
    pages = "181602",
    year = "2006"
}

@article{Sully:2020pza,
    author = "Sully, James and Van Raamsdonk, Mark and Wakeham, David",
    title = "{BCFT entanglement entropy at large central charge and the black hole interior}",
    eprint = "2004.13088",
    archivePrefix = "arXiv",
    primaryClass = "hep-th",
    doi = "10.1007/JHEP03(2021)167",
    journal = "JHEP",
    volume = "03",
    pages = "167",
    year = "2021"
}

@article{Almheiri:2019psf,
    author = "Almheiri, Ahmed and Engelhardt, Netta and Marolf, Donald and Maxfield, Henry",
    title = "{The entropy of bulk quantum fields and the entanglement wedge of an evaporating black hole}",
    eprint = "1905.08762",
    archivePrefix = "arXiv",
    primaryClass = "hep-th",
    doi = "10.1007/JHEP12(2019)063",
    journal = "JHEP",
    volume = "12",
    pages = "063",
    year = "2019"
}

@article{Ghosh:2019rcj,
    author = "Ghosh, Animik and Maxfield, Henry and Turiaci, Gustavo J.",
    title = "{A universal Schwarzian sector in two-dimensional conformal field theories}",
    eprint = "1912.07654",
    archivePrefix = "arXiv",
    primaryClass = "hep-th",
    doi = "10.1007/JHEP05(2020)104",
    journal = "JHEP",
    volume = "05",
    pages = "104",
    year = "2020"
}

@article{Almheiri:2014cka,
    author = "Almheiri, Ahmed and Polchinski, Joseph",
    title = "{Models of AdS$_{2}$ backreaction and holography}",
    eprint = "1402.6334",
    archivePrefix = "arXiv",
    primaryClass = "hep-th",
    doi = "10.1007/JHEP11(2015)014",
    journal = "JHEP",
    volume = "11",
    pages = "014",
    year = "2015"
}

@article{Jensen:2016pah,
    author = "Jensen, Kristan",
    title = "{Chaos in AdS$_2$ Holography}",
    eprint = "1605.06098",
    archivePrefix = "arXiv",
    primaryClass = "hep-th",
    doi = "10.1103/PhysRevLett.117.111601",
    journal = "Phys. Rev. Lett.",
    volume = "117",
    number = "11",
    pages = "111601",
    year = "2016"
}

@article{Maldacena:2016upp,
    author = "Maldacena, Juan and Stanford, Douglas and Yang, Zhenbin",
    title = "{Conformal symmetry and its breaking in two dimensional Nearly Anti-de-Sitter space}",
    eprint = "1606.01857",
    archivePrefix = "arXiv",
    primaryClass = "hep-th",
    doi = "10.1093/ptep/ptw124",
    journal = "PTEP",
    volume = "2016",
    number = "12",
    pages = "12C104",
    year = "2016"
}

@article{Engelsoy:2016xyb,
    author = {Engels{\"o}y, Julius and Mertens, Thomas G. and Verlinde, Herman},
    title = "{An investigation of AdS$_{2}$ backreaction and holography}",
    eprint = "1606.03438",
    archivePrefix = "arXiv",
    primaryClass = "hep-th",
    doi = "10.1007/JHEP07(2016)139",
    journal = "JHEP",
    volume = "07",
    pages = "139",
    year = "2016"
}

@article{Maldacena:2016hyu,
    author = "Maldacena, Juan and Stanford, Douglas",
    title = "{Remarks on the Sachdev-Ye-Kitaev model}",
    eprint = "1604.07818",
    archivePrefix = "arXiv",
    primaryClass = "hep-th",
    doi = "10.1103/PhysRevD.94.106002",
    journal = "Phys. Rev. D",
    volume = "94",
    number = "10",
    pages = "106002",
    year = "2016"
}

@article{Cardy:2004hm,
    author = "Cardy, John L.",
    title = "{Boundary conformal field theory}",
    eprint = "hep-th/0411189",
    archivePrefix = "arXiv",
    month = "11",
    year = "2004"
}

@article{Brown:1986nw,
    author = "Brown, J. David and Henneaux, M.",
    title = "{Central Charges in the Canonical Realization of Asymptotic Symmetries: An Example from Three-Dimensional Gravity}",
    doi = "10.1007/BF01211590",
    journal = "Commun. Math. Phys.",
    volume = "104",
    pages = "207--226",
    year = "1986"
}

@article{Perlmutter:2013paa,
    author = "Perlmutter, Eric",
    title = "{Comments on Renyi entropy in AdS$_3$/CFT$_2$}",
    eprint = "1312.5740",
    archivePrefix = "arXiv",
    primaryClass = "hep-th",
    doi = "10.1007/JHEP05(2014)052",
    journal = "JHEP",
    volume = "05",
    pages = "052",
    year = "2014"
}

@article{Hollowood:2024uuf,
    author = "Hollowood, Timothy J. and Kumar, S. Prem and Piper, Luke C.",
    title = "{Replica R{\'e}nyi wormholes and generalised modular entropy in JT gravity}",
    eprint = "2406.16339",
    archivePrefix = "arXiv",
    primaryClass = "hep-th",
    doi = "10.1007/JHEP10(2024)169",
    journal = "JHEP",
    volume = "10",
    pages = "169",
    year = "2024"
}

@article{Polchinski:2016xgd,
    author = "Polchinski, Joseph and Rosenhaus, Vladimir",
    title = "{The Spectrum in the Sachdev-Ye-Kitaev Model}",
    eprint = "1601.06768",
    archivePrefix = "arXiv",
    primaryClass = "hep-th",
    doi = "10.1007/JHEP04(2016)001",
    journal = "JHEP",
    volume = "04",
    pages = "001",
    year = "2016"
}

@article{Teitelboim:1983ux,
    author = "Teitelboim, C.",
    title = "{Gravitation and Hamiltonian Structure in Two Space-Time Dimensions}",
    doi = "10.1016/0370-2693(83)90012-6",
    journal = "Phys. Lett. B",
    volume = "126",
    pages = "41--45",
    year = "1983"
}

@article{Sachdev:1992fk,
    author = "Sachdev, Subir and Ye, Jinwu",
    title = "{Gapless spin fluid ground state in a random, quantum Heisenberg magnet}",
    eprint = "cond-mat/9212030",
    archivePrefix = "arXiv",
    reportNumber = "PRINT-93-0077",
    doi = "10.1103/PhysRevLett.70.3339",
    journal = "Phys. Rev. Lett.",
    volume = "70",
    pages = "3339",
    year = "1993"
}

@article{Jackiw:1984je,
    author = "Jackiw, R.",
    editor = "Baier, R. and Satz, H.",
    title = "{Lower Dimensional Gravity}",
    reportNumber = "MIT-CTP-1203",
    doi = "10.1016/0550-3213(85)90448-1",
    journal = "Nucl. Phys. B",
    volume = "252",
    pages = "343--356",
    year = "1985"
}

@article{Neuenfeld:2024gta,
    author = "Neuenfeld, Dominik and Svesko, Andrew and Sybesma, Watse",
    title = "{Liouville gravity at the end of the world:deformed defects in AdS/BCFT}",
    eprint = "2404.07260",
    archivePrefix = "arXiv",
    primaryClass = "hep-th",
    doi = "10.1007/JHEP07(2024)215",
    journal = "JHEP",
    volume = "07",
    pages = "215",
    year = "2024"
}

@article{Calabrese:2009qy,
    author = "Calabrese, Pasquale and Cardy, John",
    title = "{Entanglement entropy and conformal field theory}",
    eprint = "0905.4013",
    archivePrefix = "arXiv",
    primaryClass = "cond-mat.stat-mech",
    doi = "10.1088/1751-8113/42/50/504005",
    journal = "J. Phys. A",
    volume = "42",
    pages = "504005",
    year = "2009"
}

@article{Takayanagi:2011zk,
    author = "Takayanagi, Tadashi",
    title = "{Holographic Dual of BCFT}",
    eprint = "1105.5165",
    archivePrefix = "arXiv",
    primaryClass = "hep-th",
    reportNumber = "IPMU11-0091",
    doi = "10.1103/PhysRevLett.107.101602",
    journal = "Phys. Rev. Lett.",
    volume = "107",
    pages = "101602",
    year = "2011"
}

@article{Ohmori_2015,
   title={Physics at the entangling surface},
   volume={2015},
   ISSN={1742-5468},
   url={http://dx.doi.org/10.1088/1742-5468/2015/04/P04010},
   DOI={10.1088/1742-5468/2015/04/p04010},
   number={4},
   journal={Journal of Statistical Mechanics: Theory and Experiment},
   publisher={IOP Publishing},
   author={Ohmori, Kantaro and Tachikawa, Yuji},
   year={2015},
   month=apr, pages={P04010} }

@article{Aguilar-Gutierrez:2023tic,
    author = "Aguilar-Gutierrez, Sergio E. and Patra, Ayan K. and Pedraza, Juan F.",
    title = "{Entangled universes in dS wedge holography}",
    eprint = "2308.05666",
    archivePrefix = "arXiv",
    primaryClass = "hep-th",
    reportNumber = "IFT-UAM/CSIC-23-95",
    doi = "10.1007/JHEP10(2023)156",
    journal = "JHEP",
    volume = "10",
    pages = "156",
    year = "2023"
}

@article{Geng:2022slq,
    author = "Geng, Hao and Karch, Andreas and Perez-Pardavila, Carlos and Raju, Suvrat and Randall, Lisa and Riojas, Marcos and Shashi, Sanjit",
    title = "{Jackiw-Teitelboim Gravity from the Karch-Randall Braneworld}",
    eprint = "2206.04695",
    archivePrefix = "arXiv",
    primaryClass = "hep-th",
    doi = "10.1103/PhysRevLett.129.231601",
    journal = "Phys. Rev. Lett.",
    volume = "129",
    number = "23",
    pages = "231601",
    year = "2022"
}

@article{AffleckLudwig1991,
  title        = {Universal noninteger ``ground-state degeneracy'' in critical quantum systems},
  author       = {Affleck, Ian and Ludwig, Andreas W. W.},
  journal      = {Phys. Rev. Lett.},
  volume       = {67},
  pages        = {161--164},
  year         = {1991},
  doi          = {10.1103/PhysRevLett.67.161},
}

@article{Turiaci:2024cad,
    author = "Turiaci, Gustavo J.",
    title = "{Les Houches lectures on two-dimensional gravity and holography}",
    eprint = "2412.09537",
    archivePrefix = "arXiv",
    primaryClass = "hep-th",
    doi = "10.21468/SciPostPhysLectNotes.113",
    month = "12",
    year = "2024"
}

@book{schottenloher1997mathematical,
  title={A mathematical introduction to conformal field theory: Based on a series of lectures given at the Mathematisches Institut der Universitaet Hamburg},
  author={Schottenloher, Martin},
  year={1997},
  publisher={Springer}
}

@article{Hartman:2013mia,
    author = "Hartman, Thomas",
    title = "{Entanglement Entropy at Large Central Charge}",
    eprint = "1303.6955",
    archivePrefix = "arXiv",
    primaryClass = "hep-th",
    month = "3",
    year = "2013"
}

@article{Miao:2018dvm,
    author = "Miao, Rong-Xin",
    title = "{Casimir Effect, Weyl Anomaly and Displacement Operator in Boundary Conformal Field Theory}",
    eprint = "1808.05783",
    archivePrefix = "arXiv",
    primaryClass = "hep-th",
    doi = "10.1007/JHEP07(2019)098",
    journal = "JHEP",
    volume = "07",
    pages = "098",
    year = "2019"
}

@article{Billo:2016cpy,
    author = "Bill{\`o}, Marco and Gon{\c{c}}alves, Vasco and Lauria, Edoardo and Meineri, Marco",
    title = "{Defects in conformal field theory}",
    eprint = "1601.02883",
    archivePrefix = "arXiv",
    primaryClass = "hep-th",
    doi = "10.1007/JHEP04(2016)091",
    journal = "JHEP",
    volume = "04",
    pages = "091",
    year = "2016"
}

@book{blumenhagen2012basic,
  title={Basic concepts of string theory},
  author={Blumenhagen, Ralph and L{\"u}st, Dieter and Theisen, Stefan},
  year={2012},
  publisher={Springer Science \& Business Media}
}

@article{Calabrese:2004eu,
    author = "Calabrese, Pasquale and Cardy, John L.",
    title = "{Entanglement entropy and quantum field theory}",
    eprint = "hep-th/0405152",
    archivePrefix = "arXiv",
    doi = "10.1088/1742-5468/2004/06/P06002",
    journal = "J. Stat. Mech.",
    volume = "0406",
    pages = "P06002",
    year = "2004"
}

@article{Cardy:2007mb,
    author = "Cardy, J. L. and Castro-Alvaredo, O. A. and Doyon, B.",
    title = "{Form factors of branch-point twist fields in quantum integrable models and entanglement entropy}",
    eprint = "0706.3384",
    archivePrefix = "arXiv",
    primaryClass = "hep-th",
    doi = "10.1007/s10955-007-9422-x",
    journal = "J. Statist. Phys.",
    volume = "130",
    pages = "129--168",
    year = "2008"
}

@article{Akal:2021foz,
    author = "Akal, Ibrahim and Kusuki, Yuya and Shiba, Noburo and Takayanagi, Tadashi and Wei, Zixia",
    title = "{Holographic moving mirrors}",
    eprint = "2106.11179",
    archivePrefix = "arXiv",
    primaryClass = "hep-th",
    reportNumber = "YITP-21-53, IPMU21-0035, RIKEN-iTHEMS-Report-21",
    doi = "10.1088/1361-6382/ac2c1b",
    journal = "Class. Quant. Grav.",
    volume = "38",
    number = "22",
    pages = "224001",
    year = "2021"
}

@article{Ares:2022koq,
    author = "Ares, Filiberto and Murciano, Sara and Calabrese, Pasquale",
    title = "{Entanglement asymmetry as a probe of symmetry breaking}",
    eprint = "2207.14693",
    archivePrefix = "arXiv",
    primaryClass = "cond-mat.stat-mech",
    doi = "10.1038/s41467-023-37747-8",
    journal = "Nature Commun.",
    volume = "14",
    number = "1",
    pages = "2036",
    year = "2023"
}

@article{Fossati:2024ekt,
    author = "Fossati, Michele and Rylands, Colin and Calabrese, Pasquale",
    title = "{Entanglement asymmetry in CFT with boundary symmetry breaking}",
    eprint = "2411.10244",
    archivePrefix = "arXiv",
    primaryClass = "hep-th",
    doi = "10.1007/JHEP06(2025)089",
    journal = "JHEP",
    volume = "06",
    pages = "089",
    year = "2025"
}

@article{Benini:2024xjv,
    author = "Benini, Francesco and Godet, Victor and Singh, Amartya Harsh",
    title = "{Entanglement asymmetry in conformal field theory and holography}",
    eprint = "2407.07969",
    archivePrefix = "arXiv",
    primaryClass = "hep-th",
    reportNumber = "SISSA 14/2024/FISI",
    doi = "10.1093/ptep/ptaf080",
    journal = "PTEP",
    volume = "2025",
    pages = "6",
    year = "2025"
}

@article{Mertens:2018fds,
    author = "Mertens, Thomas G.",
    title = "{The Schwarzian theory {\textemdash} origins}",
    eprint = "1801.09605",
    archivePrefix = "arXiv",
    primaryClass = "hep-th",
    doi = "10.1007/JHEP05(2018)036",
    journal = "JHEP",
    volume = "05",
    pages = "036",
    year = "2018"
}

@article{Lam:2018pvp,
    author = "Lam, Ho Tat and Mertens, Thomas G. and Turiaci, Gustavo J. and Verlinde, Herman",
    title = "{Shockwave S-matrix from Schwarzian Quantum Mechanics}",
    eprint = "1804.09834",
    archivePrefix = "arXiv",
    primaryClass = "hep-th",
    doi = "10.1007/JHEP11(2018)182",
    journal = "JHEP",
    volume = "11",
    pages = "182",
    year = "2018"
}

@article{Saad:2019lba,
    author = "Saad, Phil and Shenker, Stephen H. and Stanford, Douglas",
    title = "{JT gravity as a matrix integral}",
    eprint = "1903.11115",
    archivePrefix = "arXiv",
    primaryClass = "hep-th",
    month = "3",
    year = "2019"
}

@article{Suzuki:2021zbe,
    author = "Suzuki, Kenta and Takayanagi, Tadashi",
    title = "{JT gravity limit of Liouville CFT and matrix model}",
    eprint = "2108.12096",
    archivePrefix = "arXiv",
    primaryClass = "hep-th",
    reportNumber = "YITP-21-88, IPMU21-0054",
    doi = "10.1007/JHEP11(2021)137",
    journal = "JHEP",
    volume = "11",
    pages = "137",
    year = "2021"
}

@article{Penington:2019npb,
    author = "Penington, Geoffrey",
    title = "{Entanglement Wedge Reconstruction and the Information Paradox}",
    eprint = "1905.08255",
    archivePrefix = "arXiv",
    primaryClass = "hep-th",
    doi = "10.1007/JHEP09(2020)002",
    journal = "JHEP",
    volume = "09",
    pages = "002",
    year = "2020"
}

@article{Bousso:2015mna,
    author = "Bousso, Raphael and Fisher, Zachary and Leichenauer, Stefan and Wall, Aron C.",
    title = "{Quantum focusing conjecture}",
    eprint = "1506.02669",
    archivePrefix = "arXiv",
    primaryClass = "hep-th",
    doi = "10.1103/PhysRevD.93.064044",
    journal = "Phys. Rev. D",
    volume = "93",
    number = "6",
    pages = "064044",
    year = "2016"
}

@article{Karch:2000ct,
    author = "Karch, Andreas and Randall, Lisa",
    editor = "Duff, Michael J. and Liu, J. T. and Lu, J.",
    title = "{Locally localized gravity}",
    eprint = "hep-th/0011156",
    archivePrefix = "arXiv",
    reportNumber = "MIT-CTP-3099",
    doi = "10.1088/1126-6708/2001/05/008",
    journal = "JHEP",
    volume = "05",
    pages = "008",
    year = "2001"
}

@article{Callebaut:2018xfu,
    author = "Callebaut, Nele",
    title = "{The gravitational dynamics of kinematic space}",
    eprint = "1808.10431",
    archivePrefix = "arXiv",
    primaryClass = "hep-th",
    doi = "10.1007/JHEP02(2019)153",
    journal = "JHEP",
    volume = "02",
    pages = "153",
    year = "2019"
}

@article{Callebaut:2018nlq,
    author = "Callebaut, Nele and Verlinde, Herman",
    title = "{Entanglement Dynamics in 2D CFT with Boundary: Entropic origin of JT gravity and Schwarzian QM}",
    eprint = "1808.05583",
    archivePrefix = "arXiv",
    primaryClass = "hep-th",
    doi = "10.1007/JHEP05(2019)045",
    journal = "JHEP",
    volume = "05",
    pages = "045",
    year = "2019"
}

@inproceedings{Strominger:1994tn,
    author = "Strominger, Andrew",
    title = "{Les Houches lectures on black holes}",
    booktitle = "{NATO Advanced Study Institute: Les Houches Summer School, Session 62: Fluctuating Geometries in Statistical Mechanics and Field Theory}",
    eprint = "hep-th/9501071",
    archivePrefix = "arXiv",
    month = "8",
    year = "1994"
}

@article{Moitra:2019bub,
    author = "Moitra, Upamanyu and Sake, Sunil Kumar and Trivedi, Sandip P. and Vishal, V.",
    title = "{Jackiw-Teitelboim Gravity and Rotating Black Holes}",
    eprint = "1905.10378",
    archivePrefix = "arXiv",
    primaryClass = "hep-th",
    reportNumber = "TIFR/TH/19-17",
    doi = "10.1007/JHEP11(2019)047",
    journal = "JHEP",
    volume = "11",
    pages = "047",
    year = "2019"
}

@article{Nayak:2018qej,
    author = "Nayak, Pranjal and Shukla, Ashish and Soni, Ronak M. and Trivedi, Sandip P. and Vishal, V.",
    title = "{On the Dynamics of Near-Extremal Black Holes}",
    eprint = "1802.09547",
    archivePrefix = "arXiv",
    primaryClass = "hep-th",
    reportNumber = "TIFR/TH/17-35, TIFR-TH-17-35",
    doi = "10.1007/JHEP09(2018)048",
    journal = "JHEP",
    volume = "09",
    pages = "048",
    year = "2018"
}

@article{Almheiri:2016fws,
    author = "Almheiri, Ahmed and Kang, Byungwoo",
    title = "{Conformal Symmetry Breaking and Thermodynamics of Near-Extremal Black Holes}",
    eprint = "1606.04108",
    archivePrefix = "arXiv",
    primaryClass = "hep-th",
    doi = "10.1007/JHEP10(2016)052",
    journal = "JHEP",
    volume = "10",
    pages = "052",
    year = "2016"
}

@article{Stanford:2019vob,
    author = "Stanford, Douglas and Witten, Edward",
    title = "{JT gravity and the ensembles of random matrix theory}",
    eprint = "1907.03363",
    archivePrefix = "arXiv",
    primaryClass = "hep-th",
    doi = "10.4310/ATMP.2020.v24.n6.a4",
    journal = "Adv. Theor. Math. Phys.",
    volume = "24",
    number = "6",
    pages = "1475--1680",
    year = "2020"
}

@article{Rozali:2019day,
    author = "Rozali, Moshe and Sully, James and Van Raamsdonk, Mark and Waddell, Christopher and Wakeham, David",
    title = "{Information radiation in BCFT models of black holes}",
    eprint = "1910.12836",
    archivePrefix = "arXiv",
    primaryClass = "hep-th",
    doi = "10.1007/JHEP05(2020)004",
    journal = "JHEP",
    volume = "05",
    pages = "004",
    year = "2020"
}

@article{Almheiri:2019hni,
    author = "Almheiri, Ahmed and Mahajan, Raghu and Maldacena, Juan and Zhao, Ying",
    title = "{The Page curve of Hawking radiation from semiclassical geometry}",
    eprint = "1908.10996",
    archivePrefix = "arXiv",
    primaryClass = "hep-th",
    doi = "10.1007/JHEP03(2020)149",
    journal = "JHEP",
    volume = "03",
    pages = "149",
    year = "2020"
}

@article{Callebaut:2025thw,
    author = "Callebaut, Nele and Selle, Matteo",
    title = "{Setting T$^{2}$ free for braneworld holography}",
    eprint = "2510.01099",
    archivePrefix = "arXiv",
    primaryClass = "hep-th",
    doi = "10.1007/JHEP02(2026)234",
    journal = "JHEP",
    volume = "02",
    pages = "234",
    year = "2026"
}

@article{Randall:1999vf,
    author = "Randall, Lisa and Sundrum, Raman",
    title = "{An Alternative to compactification}",
    eprint = "hep-th/9906064",
    archivePrefix = "arXiv",
    reportNumber = "MIT-CTP-2874, PUPT-1867, BUHEP-99-13",
    doi = "10.1103/PhysRevLett.83.4690",
    journal = "Phys. Rev. Lett.",
    volume = "83",
    pages = "4690--4693",
    year = "1999"
}

@article{Geng:2020qvw,
    author = "Geng, Hao and Karch, Andreas",
    title = "{Massive islands}",
    eprint = "2006.02438",
    archivePrefix = "arXiv",
    primaryClass = "hep-th",
    doi = "10.1007/JHEP09(2020)121",
    journal = "JHEP",
    volume = "09",
    pages = "121",
    year = "2020"
}

@article{Chen:2020uac,
    author = "Chen, Hong Zhe and Myers, Robert C. and Neuenfeld, Dominik and Reyes, Ignacio A. and Sandor, Joshua",
    title = "{Quantum Extremal Islands Made Easy, Part I: Entanglement on the Brane}",
    eprint = "2006.04851",
    archivePrefix = "arXiv",
    primaryClass = "hep-th",
    doi = "10.1007/JHEP10(2020)166",
    journal = "JHEP",
    volume = "10",
    pages = "166",
    year = "2020"
}

@article{Meineri:2019ycm,
    author = "Meineri, Marco and Penedones, Joao and Rousset, Antonin",
    title = "{Colliders and conformal interfaces}",
    eprint = "1904.10974",
    archivePrefix = "arXiv",
    primaryClass = "hep-th",
    doi = "10.1007/JHEP02(2020)138",
    journal = "JHEP",
    volume = "02",
    pages = "138",
    year = "2020"
}

@article{Karch:2025hof,
    author = "Karch, Andreas and Youssef, Merna",
    title = "{Dissipation in open holography}",
    eprint = "2509.14312",
    archivePrefix = "arXiv",
    primaryClass = "hep-th",
    reportNumber = "UT-WI-30-2025",
    doi = "10.1007/JHEP12(2025)157",
    journal = "JHEP",
    volume = "12",
    pages = "157",
    year = "2025"
}

@article{Geng:2020fxl,
    author = "Geng, Hao and Karch, Andreas and Perez-Pardavila, Carlos and Raju, Suvrat and Randall, Lisa and Riojas, Marcos and Shashi, Sanjit",
    title = "{Information Transfer with a Gravitating Bath}",
    eprint = "2012.04671",
    archivePrefix = "arXiv",
    primaryClass = "hep-th",
    doi = "10.21468/SciPostPhys.10.5.103",
    journal = "SciPost Phys.",
    volume = "10",
    number = "5",
    pages = "103",
    year = "2021"
}

@article{Iliesiu:2020qvm,
    author = "Iliesiu, Luca V. and Turiaci, Gustavo J.",
    title = "{The statistical mechanics of near-extremal black holes}",
    eprint = "2003.02860",
    archivePrefix = "arXiv",
    primaryClass = "hep-th",
    doi = "10.1007/JHEP05(2021)145",
    journal = "JHEP",
    volume = "05",
    pages = "145",
    year = "2021"
}

@article{York:1972sj,
    author = "York, Jr., James W.",
    title = "{Role of conformal three geometry in the dynamics of gravitation}",
    doi = "10.1103/PhysRevLett.28.1082",
    journal = "Phys. Rev. Lett.",
    volume = "28",
    pages = "1082--1085",
    year = "1972"
}

@article{Gibbons:1976ue,
    author = "Gibbons, G. W. and Hawking, S. W.",
    title = "{Action Integrals and Partition Functions in Quantum Gravity}",
    reportNumber = "PRINT-76-0995 (CAMBRIDGE)",
    doi = "10.1103/PhysRevD.15.2752",
    journal = "Phys. Rev. D",
    volume = "15",
    pages = "2752--2756",
    year = "1977"
}

@article{Susskind:1994sm,
    author = "Susskind, Leonard and Uglum, John",
    title = "{Black hole entropy in canonical quantum gravity and superstring theory}",
    eprint = "hep-th/9401070",
    archivePrefix = "arXiv",
    reportNumber = "SU-ITP-94-1",
    doi = "10.1103/PhysRevD.50.2700",
    journal = "Phys. Rev. D",
    volume = "50",
    pages = "2700--2711",
    year = "1994"
}

@article{Engelhardt:2014gca,
    author = "Engelhardt, Netta and Wall, Aron C.",
    title = "{Quantum Extremal Surfaces: Holographic Entanglement Entropy beyond the Classical Regime}",
    eprint = "1408.3203",
    archivePrefix = "arXiv",
    primaryClass = "hep-th",
    doi = "10.1007/JHEP01(2015)073",
    journal = "JHEP",
    volume = "01",
    pages = "073",
    year = "2015"
}

@article{Bousso:2022hlz,
    author = "Bousso, Raphael and Penington, Geoff",
    title = "{Entanglement wedges for gravitating regions}",
    eprint = "2208.04993",
    archivePrefix = "arXiv",
    primaryClass = "hep-th",
    doi = "10.1103/PhysRevD.107.086002",
    journal = "Phys. Rev. D",
    volume = "107",
    number = "8",
    pages = "086002",
    year = "2023"
}

@article{Calabrese:2009ez,
    author = "Calabrese, Pasquale and Cardy, John and Tonni, Erik",
    title = "{Entanglement entropy of two disjoint intervals in conformal field theory}",
    eprint = "0905.2069",
    archivePrefix = "arXiv",
    primaryClass = "hep-th",
    doi = "10.1088/1742-5468/2009/11/P11001",
    journal = "J. Stat. Mech.",
    volume = "0911",
    pages = "P11001",
    year = "2009"
}

@article{Calabrese:2010he,
    author = "Calabrese, Pasquale and Cardy, John and Tonni, Erik",
    title = "{Entanglement entropy of two disjoint intervals in conformal field theory II}",
    eprint = "1011.5482",
    archivePrefix = "arXiv",
    primaryClass = "hep-th",
    reportNumber = "NSF-KITP-10-152, MIT-CTP 4194",
    doi = "10.1088/1742-5468/2011/01/P01021",
    journal = "J. Stat. Mech.",
    volume = "1101",
    pages = "P01021",
    year = "2011"
}

@article{Belavin:1984vu,
    author = "Belavin, A. A. and Polyakov, Alexander M. and Zamolodchikov, A. B.",
    editor = "Khalatnikov, I. M. and Mineev, V. P.",
    title = "{Infinite Conformal Symmetry in Two-Dimensional Quantum Field Theory}",
    reportNumber = "CERN-TH-3827",
    doi = "10.1016/0550-3213(84)90052-X",
    journal = "Nucl. Phys. B",
    volume = "241",
    pages = "333--380",
    year = "1984"
}

@article{Zamolodchikov:1987avt,
    author = "Zamolodchikov, Al. B.",
    title = "{Conformal symmetry in two-dimensional space: Recursion representation of conformal block}",
    doi = "10.1007/BF01022967",
    journal = "Theor. Math. Phys.",
    volume = "73",
    number = "1",
    pages = "1088--1093",
    year = "1987"
}

@article{Numasawa:2022cni,
    author = "Numasawa, Tokiro and Tsiares, Ioannis",
    title = "{Universal dynamics of heavy operators in boundary CFT$_{2}$}",
    eprint = "2202.01633",
    archivePrefix = "arXiv",
    primaryClass = "hep-th",
    doi = "10.1007/JHEP08(2022)156",
    journal = "JHEP",
    volume = "08",
    pages = "156",
    year = "2022"
}

@article{Hartman:2014oaa,
    author = "Hartman, Thomas and Keller, Christoph A. and Stoica, Bogdan",
    title = "{Universal Spectrum of 2d Conformal Field Theory in the Large c Limit}",
    eprint = "1405.5137",
    archivePrefix = "arXiv",
    primaryClass = "hep-th",
    reportNumber = "CALT-68-2889, RUNHETC-2014-07",
    doi = "10.1007/JHEP09(2014)118",
    journal = "JHEP",
    volume = "09",
    pages = "118",
    year = "2014"
}

@article{Pappadopulo:2012jk,
    author = "Pappadopulo, Duccio and Rychkov, Slava and Espin, Johnny and Rattazzi, Riccardo",
    title = "{OPE Convergence in Conformal Field Theory}",
    eprint = "1208.6449",
    archivePrefix = "arXiv",
    primaryClass = "hep-th",
    reportNumber = "LPTENS-12-31",
    doi = "10.1103/PhysRevD.86.105043",
    journal = "Phys. Rev. D",
    volume = "86",
    pages = "105043",
    year = "2012"
}

@article{QDF,
    author = "Llorens Giralt, Quim and Neuenfeld, Dominik and Taube, Frederik",
    title = "{Interactions between Shocks and Spacetime Boundaries (to appear)}",
    eprint = "260X.XXXX",
    archivePrefix = "arXiv",
}

@article{Perlmutter:2015iya,
    author = "Perlmutter, Eric",
    title = "{Virasoro conformal blocks in closed form}",
    eprint = "1502.07742",
    archivePrefix = "arXiv",
    primaryClass = "hep-th",
    doi = "10.1007/JHEP08(2015)088",
    journal = "JHEP",
    volume = "08",
    pages = "088",
    year = "2015"
}

@article{Suzuki:2022xwv,
    author = "Suzuki, Kenta and Takayanagi, Tadashi",
    title = "{BCFT and Islands in two dimensions}",
    eprint = "2202.08462",
    archivePrefix = "arXiv",
    primaryClass = "hep-th",
    reportNumber = "YITP-22-14, IPMU22-0002",
    doi = "10.1007/JHEP06(2022)095",
    journal = "JHEP",
    volume = "06",
    pages = "095",
    year = "2022"
}

\end{document}